\documentclass[journal]{IEEEtran}
\ifCLASSINFOpdf
\else
\fi
\usepackage{graphics} 
\usepackage{epsfig} 
\usepackage{times} 
\usepackage{amsmath} 
\usepackage{amssymb}  
\usepackage[svgnames]{xcolor}
\usepackage{amsfonts}

\usepackage{algorithm}
\usepackage{algpseudocode}

\usepackage{setspace}
\usepackage{placeins}
\usepackage{subcaption}
\usepackage{booktabs}
\usepackage{multirow}
\usepackage{url}
\usepackage[shortcuts]{extdash}

\usepackage{tikz}
\usetikzlibrary{shapes.misc, positioning}
\usepackage{varwidth}
\usepackage{ragged2e}
\usepackage{subcaption}
\usepackage{gensymb}

\usepackage{comment}
\usepackage{tabularx}

\newcommand{\rv}[1]{\textcolor{black}{#1}}
\newcommand{\rrv}[1]{\textcolor{black}{#1}}

\begin{document}
%
\title{Listening for Sirens: Locating and Classifying Acoustic Alarms in City Scenes}
%
%
%

\author{Letizia Marchegiani~\IEEEmembership{Member,~IEEE,} and Paul Newman~\IEEEmembership{Fellow,~IEEE}
\thanks{L. Marchegiani is with the Department of Electronic Systems, Aalborg University, 9200 Aalborg \O, Denmark (email: lm@es.aau.dk). 

P. Newman is with the Oxford Robotics Institute, Department of Engineering Science, University of Oxford, Oxford OX1 3PJ, U.K. (email: pnewman@robots.ox.ac.uk)}}

%
%

\markboth{IEEE Transactions on Intelligent Transportation Systems . PREPRINT VERSION. ACCEPTED MARCH, 2022}%
{Shell \MakeLowercase{\textit{et al.}}: Bare Demo of IEEEtran.cls for IEEE Journals}
%



\maketitle

\begin{abstract}
This paper is about acoustic event detection and sound source localisation in  urban scenarios. Specifically, we are interested in detecting and localising horns and sirens of emergency vehicles. Urban scenarios, though, can be characterised by copious, unstructured and unpredictable traffic noise, which can severely compromise the performance and effectiveness of traditional filtering techniques. By analysing the spectrograms of incoming stereo signals as images, we can leverage image processing techniques and obtain a demonstrably robust system. 
Indeed, image processing methods, such as convolutional neural networks, which do not operate locally, offer interesting mechanisms for background foreground separation. When applied to spectrograms, those mechanisms allow using the entire context of the soundscape to discover and learn correlations both in the time and frequency domains, \textit{de facto} implementing noise detection through semantic segmentation. In a multi-task learning scheme, together with signal denoising, we perform acoustic event classification to identify the nature of the alerting sound. Lastly, we use the denoised signals to localise the acoustic source on the ground plane, by regressing the direction of arrival of the sound. Our experimental evaluation shows an average classification rate of 94\%, and a median absolute error on the localisation of 7.5$\degree$ when operating on audio frames of 0.5 s, and of 2.5$\degree$ when operating on frames of 2.5 s. The system offers excellent performance in particularly challenging scenarios, where the noise level is remarkably high.
\end{abstract}


%
\IEEEpeerreviewmaketitle

\section{INTRODUCTION}
\label{sec:intro}
\IEEEPARstart{O}{ur} autonomous vehicles / cars are largely deaf. They typically make little if any use of auditory inputs and this paper starts to address that shortcoming. Here we approach the problem of using auditory perception in intelligent transportation to spot the presence of, and localise ``alerting events'' which carry crucial information to enable safe navigation in urban areas, and which, in some cases (\textit{e.g.} a car honking) could not be perceived by different sensing means. Specifically, we aim to detect and recognise anomalous sounds, such as car horns and sirens of emergency vehicles, and localise the respective acoustic sources. Autonomous vehicles would clearly benefit from the ability to identify and interpret those signals. An emergency vehicle approaching an intersection could be detectable long before it reaches the crossing point and despite occlusions. The possibility of having advance information of this kind  would considerably increase the time  frame allowed for a safe response from the driver, as well as the probability of successfully implementing an emergency plan for an autonomous car \cite{tatoglu2019self} (\textit{e.g.} park aside to let the emergency vehicle pass); in a smart vehicle working in semi-autonomous ``L3'' regime, it could also be used to trigger manual intervention. Furthermore, people with hearing impairments are potentially more prone
to accidents which could be avoided if these cues could be perceived \cite{hersh2010assistive}.


One of the greatest challenges in the identification of  auditory events lies in the copious and unstructured traffic noise which characterises the acoustic urban scene, and, against which, filtering techniques, traditionally used in signal processing literature, struggle to perform well. In one of our previous works \cite{marchegiani2017leveraging} we introduced the idea of treating tempo-spectral representations of the incoming audio signals (\textit{e.g.} spectrograms) as images, applying segmentation techniques for signal denoising. Identification of the various sound types was, then, performed on the filtered \emph{clean} signals, obtaining a high level of accuracy. Further analysis of the behaviour of the system when operating on the gammatonegrams of the unclean signals is also carried out, yielding to a much lower identification rate. This direct comparison serves as proof of the efficacy of the segmentation as a denoising method. 

Building on the analysis and the results of \cite{marchegiani2017leveraging},  we extend that contribution in \rv{three} directions. Firstly, rather than relying on the use of \emph{k-means} to perform image segmentation, we now employ deep learning, in the form of a \emph{U-Net} architecture \cite{long2015fully}\cite{ronneberger2015u} in a multi-task learning scheme, to simultaneously identify the nature of the acoustic event, and extract the corresponding target signal from the background noise. The conspicuous advantage that this kind of approach offers over other filtering techniques is its extreme flexibility. Traditional signal processing methods need to estimate the behaviour and the characteristics of the background noise to be able to discard it, which might be extremely challenging in urban scenarios, where the traffic noise is of a diverse nature, does not present a clear structure, and the geometry of the sound sources is unknown and unpredictable. The use of the \emph{U-Net} makes noise modelling unnecessary, as it attempts to retrieve the target signals directly. Furthermore, this method overcomes the limitations and constraints of \cite{marchegiani2017leveraging}, where the segmentation procedure was purely based on the energy characterising the stimuli in the audio mixture, and assumptions on their relationships were required. A U-Net architecture, as described in more details in Section \ref{subsec:mtl}, builds on the use of Convolutional Neural Networks (CNNs). CNNs offer the possibility to analyse the full soundscape fingerprint, and can be trained to discover time-frequency correlations. Those correlations allow to learn the characteristics of the target signal, independently of the competing noise, overcoming the limitations suffered by traditional filtering methods. We are now able to robustly address scenarios where the noise is particularly powerful compared to the target signal, and where earlier work was having major difficulties in recovering the shape of the sound of interest. Additionally, thanks to the multi-task learning scheme, the segmentation, and the consequent signal extraction, are now tailored to the class of the signal analysed. 

Secondly, \rv{while in \cite{marchegiani2017leveraging} sound source localisation is not directly addressed, in this work, we exploit} the potential of CNNs for time-frequency pattern retrieval, \rv{and} utilise a CNN-based architecture  to localise the acoustic source. Specifically, from a stereo combination (\textit{i.e.} as perceived by two different microphones, separated in space) of the recovered target signals, we estimate the direction of arrival (DoA) of the sound on the horizon plane, also known as \emph{Horizontal Localisation (HL)}. Horizontal sound source localisation techniques rely on the analysis of two main auditory cues to establish differences between the two signals in the stereo combination \cite{argentieri2015survey}. Those differences are expressed as the \emph{Interaural Time Difference (ITD)} and the \emph{Interaural Level Difference (ILD)}. The former refers to the difference in the time necessary for the acoustic wave to reach the two channels. The latter refers to the difference in the intensity of the signals in the two channels. For this kind of analysis to be successful, the sound should not be corrupted by noise. Intuitively, the cleaner the two signals are, the more accurate the resulting localisation will be; which is why spectrogram segmentation plays such a crucial role in this context. Nevertheless, small inaccuracies in the segmentation can lead to further inaccuracies in the estimation of the ITD and the ILD, which, in turn, can lead to important errors in the computation of the sound source position. In order to avoid this risk, rather than recovering the DoA of the sound from the extracted signals, we opt for learning a direct, and more robust to interference, mapping between those and the location of the sound source, through the use of CNNs. 

\rv{Thirdly, as noise represents a really challenging constraint when dealing with auditory perception in urban scenes (\textit{e.g.} \cite{parry2020pressure}), we here perform a thorough analysis of the behaviour of our system at different SNR levels.}

\section{RELATED WORK}
\label{sec:rw}
The literature reports few attempts to detect siren, and more generally, alerting urban sounds  \cite{fazenda2009acoustic}\cite{meucci2008real}\cite{schroder2013automatic}\cite{ntalampiras2009adaptive}\cite{Salamon:UrbanSound:ACMMM:14}\cite{carmel2017detection}\rv{\cite{banerjee2013participatory}}. The least recent of those attempts follow two main strategies to spot the presence of the sound of interest in the acoustic scene: they either model the characteristics of the background noise \cite{fazenda2009acoustic}\cite{ntalampiras2009adaptive}, or the ones of the target signal \cite{meucci2008real}\cite{schroder2013automatic}\rv{\cite{ebizuka2019detecting}}. In the case of the former, adaptive filtering techniques (\textit{e.g.} \cite{widrow1985adaptive}) are applied. In the case of the latter, \rv{peak searching and spectral analysis, respectively, are employed to detect the siren in the background noise}. In most recent works (\cite{schroder2013automatic} \cite{Salamon:UrbanSound:ACMMM:14}\cite{carmel2017detection}\cite{anacur2019detecting}\rv{\cite{fatimah2020automatic}\cite{banerjee2013participatory}}), instead, \rv{alarms} are detected through more traditional machine learning paradigms. Our work is closer in spirit to those, as it aims to learn the characteristics of the sound of interest independently of the nature and the features of any kind of noise potentially present. Yet, with respect the aforementioned studies, we are able to successfully address extremely challenging scenarios characterised by a remarkably low Signal-to-Noise Ratio (SNR)~($-40dB \leq SNR \leq 10 dB$). Scenarios which in previous studies, instead, are either not directly examined \cite{Salamon:UrbanSound:ACMMM:14}\cite{meucci2008real}\cite{anacur2019detecting}\rv{\cite{fatimah2020automatic}\cite{ebizuka2019detecting}\cite{banerjee2013participatory}} or yield to a substantial degrade in the classification performance \cite{schroder2013automatic}, even at a relatively high SNR ($SNR = -5 \ dB$). In \cite{carmel2017detection}, the authors intend to detect alarming sounds in different noisy scenarios, but the SNR of those scenarios \rrv{is} not mentioned.

Semantic segmentation has been widely investigated both in robotics and computer vision literature. Recent advancements in deep learning \cite{badrinarayanan2015segnet}\cite{papandreou2015weakly}\cite{long2015fully}\cite{ronneberger2015u}, especially, have determined tremendous improvements in the performance of segmentation algorithms in a wide range of application domains. Audio analysis and understanding, on the other hand, has only partially benefited from such improvements in image processing. Indeed, the idea of exploiting image processing techniques operating on tempo-spectral representations of acoustic signals (\textit{e.g.} spectrograms) is still in its infancy, with only few works (\textit{e.g.} \cite{deshpande2001classification}\cite{dutta2007text}\cite{towhid2017spectrogram}) exploring this path for speaker, birds, and music classification purposes. In \cite{marchegiani2017leveraging}, we investigated the possibility of utilising image segmentation as a noise cancelling method, obtaining promising results. We now enhance that approach, which relied on the use of \emph{k-means} \cite{fu1981survey}, by employing a more powerful segmentation method, based on a \emph{Unet} architecture \cite{long2015fully}\cite{ronneberger2015u}. As already discussed in Section~\ref{sec:intro}, this allows us to extract the sound of interest from a noisy background, even when this noise is extremely powerful, SNR  at $-40dB$, where \cite{marchegiani2017leveraging} was not able to work properly with SNRs lower than $-15dB$. Lately, \cite{mulimani2019segmentation} proposed the use of sub-spectrogram segmentation for the classification of acoustic events in domestic environments (\textit{e.g.} bag opening, dishwasher, flatware sorting). Yet, the authors do not apply any image processing technique to obtain the segmentation, but rather Singular Value Decomposition (SVD) to discriminate between low and high-energy spectral components.

Sound source localisation in robotics has been mainly concerned with human-robot interaction applications. Recently, more traditional geometry-based methods \cite{rudzyn2007real}\cite{marchegiani2009multimodal}, have been replaced by deep learning techniques \cite{he2017deep}\cite{ma2017exploiting}\cite{tse2019no}\cite{vera2018towards}\cite{he2019adaptation}\cite{zhang2018new}. In \cite{he2017deep} and \cite{ma2017exploiting}, the authors employ cross-correlation information to model the sound source location, while \cite{tse2019no} and \cite{vera2018towards} propose \emph{end-to-end} localisation. Different perspectives are analysed in \cite{he2019adaptation} and \cite{zhuang2008feature}, where the former discusses the possibility of relying on domain adaptation techniques to exploit models built in simulation in real-world acoustic scenes, and the latter introduces the concept of regional localisation. This work is close in spirit to all those studies, and, in particular to \cite{he2017deep}\cite{ma2017exploiting}. Yet, with respect to those, which focus on indoor environments and multi-speaker localisation, and the noise is either low ($SNR \geq -5dB$) or structured in the form of competing speakers, we analyse and are able to cope with outdoor scenarios characterised by the presence of a variety of unknown sources of noise of a different nature, and where the SNR can be significantly lower. \rv{Outdoor sound source localisation has been analysed in the past mainly for underwater vehicle navigation (\textit{e.g.} \cite{huang2016autonomous, rypkema2018closed}). Only few investigations have been carried out in urban scenarios (\textit{e.g.} \cite{mennitt2010multiple}), mainly focusing on the analysis of sound propagation in cluttered environments, whose map is known, and where urban noise is not present. Sound source localisation in wide-range  outdoor environments employing wireless sensor networks has been approached in \cite{faraji2019sound}. Yet, also in this case, environmental noise is not present (\textit{i.e.} experimental evaluation is carried out at $34dB$).} Outdoor scenarios and lower SNRs are investigated in \cite{tse2019no}, but, in that instance, the system is tuned to work only on a known target speaker.

\section{TECHNICAL APPROACH}
\label{sec:methods}

In this paper, we employ two different learning schemes to detect the acoustic events and localise the respective sound sources. A full representation of the framework is provided in Fig.~\ref{fig:architecture}.  The network we use to perform segmentation and event classification is reported in the blue area of the figure (\textit{cf.} Sound event Classification, \emph{SeC}), while the one used to regress the direction of arrival of the sound is reported in the pink area (\textit{cf.} Source Localisation, \emph{SL}). The acoustic events we are interested in analysing are sirens of emergency vehicles and horns. Specifically, we consider three types of sirens: ``yelp'', ``wail'' and ``hi-low'' \cite{ding2004adaptive}\cite{wagner2000guide}, which are the ones adopted in most countries. A description of the features employed is presented in Section \ref{subsec:features}, while a detailed illustration of the deep architectures used is delineated in Sections \ref{subsec:mtl} and \ref{subsec:ssl}. 

\begin{figure}[t!]
\centering
\input{unet.tkz}
\caption{The figure illustrates the entire framework, based on a two-stage approach, indicated by the blue (Sound event Classification, \emph{SeC}) and pink (Source Localisation, \emph{SL}) areas. The gammatonegrams (\textit{cf.} Section \ref{subsec:features}) of the incoming stereo sound are fed to a multi-task learning architecture (\emph{SeC}), where both segmentation and event classification are applied. The resulting segmented gammatonegrams are, then, used as masks against the original noisy ones, channel by channel (darker pink area of the \emph{SL} architecture). Lastly, the cross-correlation of the clean gammatonegrams is computed and employed to regress the direction of arrival of the sound (\emph{DoA}).
}
\label{fig:architecture}
\vspace{-6mm}
\end{figure}

\begin{figure*}%
\centering
\begin{subfigure}{.51\columnwidth}
\includegraphics[width=\columnwidth]{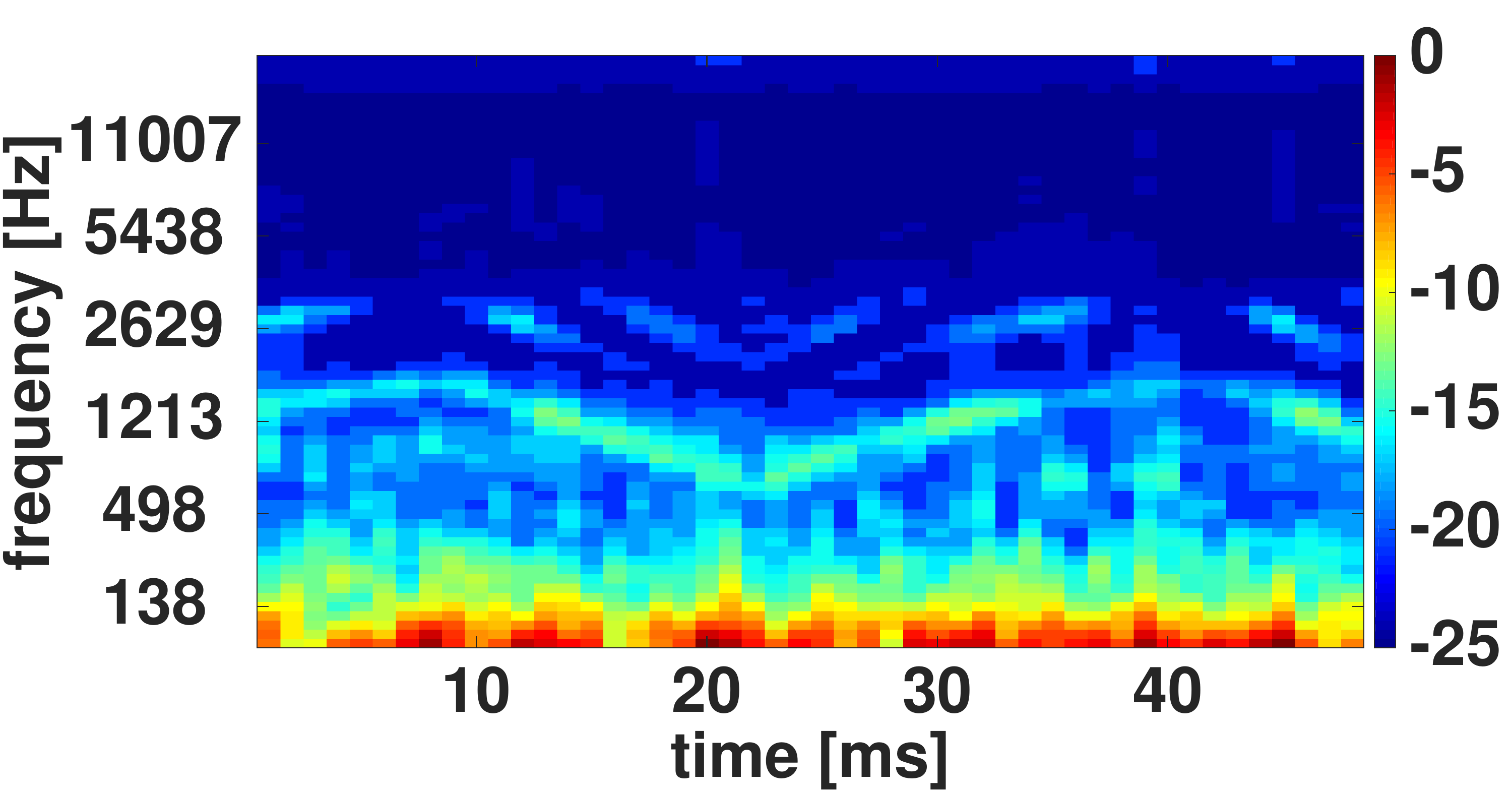}%
\caption{Siren - Yelp}%
\label{subfig:gamma_s_y}%
\end{subfigure}\hfill%
\begin{subfigure}{.51\columnwidth}
\includegraphics[width=\columnwidth]{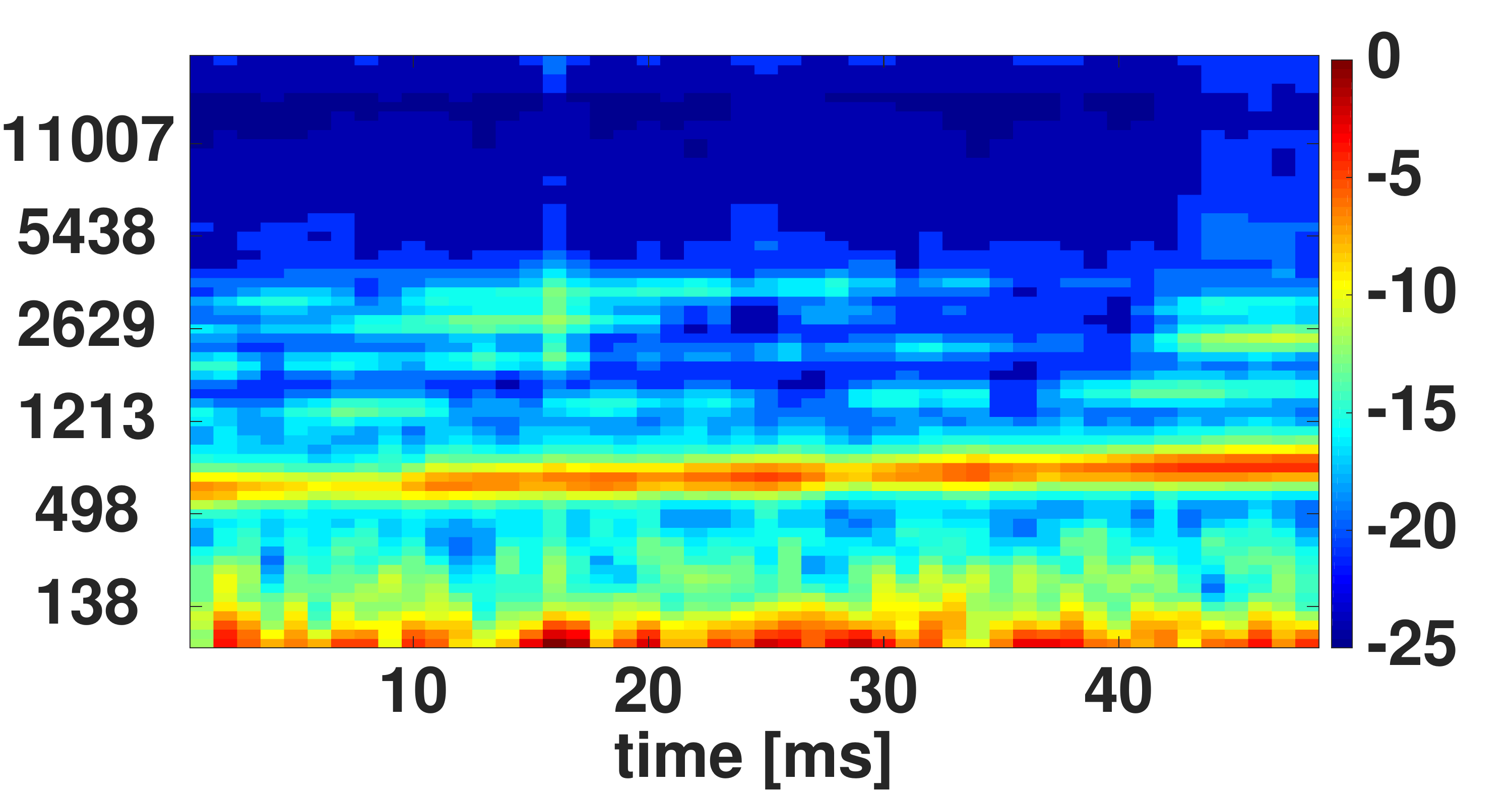}%
\caption{Siren - Wail}%
\label{subfig:gamma_s_w}%
\end{subfigure}\hfill%
\begin{subfigure}{.51\columnwidth}
\includegraphics[width=\columnwidth]{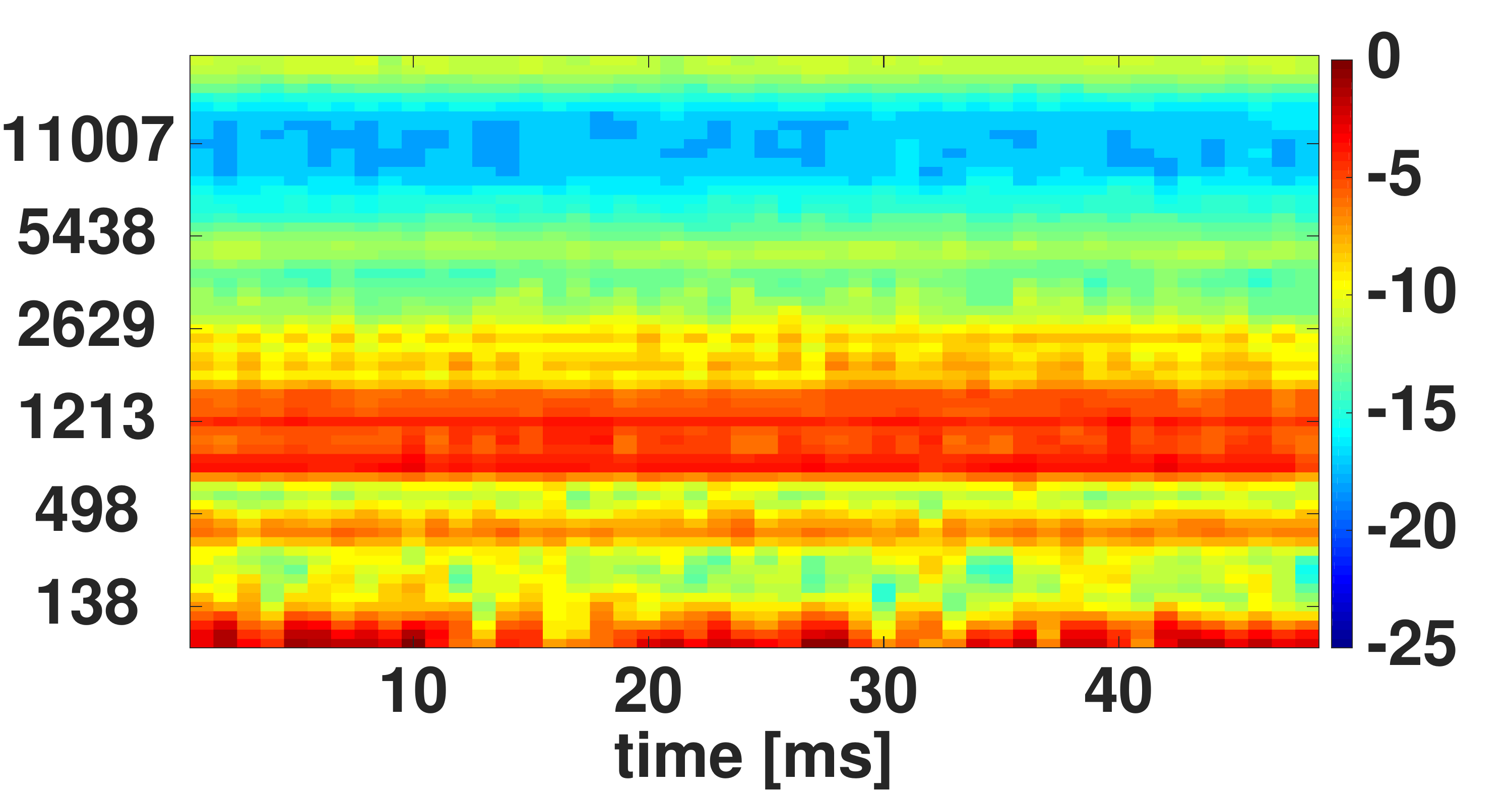}%
\caption{Horn}%
\label{subfig:gamma_h}%
\end{subfigure}\hfill%
\begin{subfigure}{.51\columnwidth}
\includegraphics[width=\columnwidth]{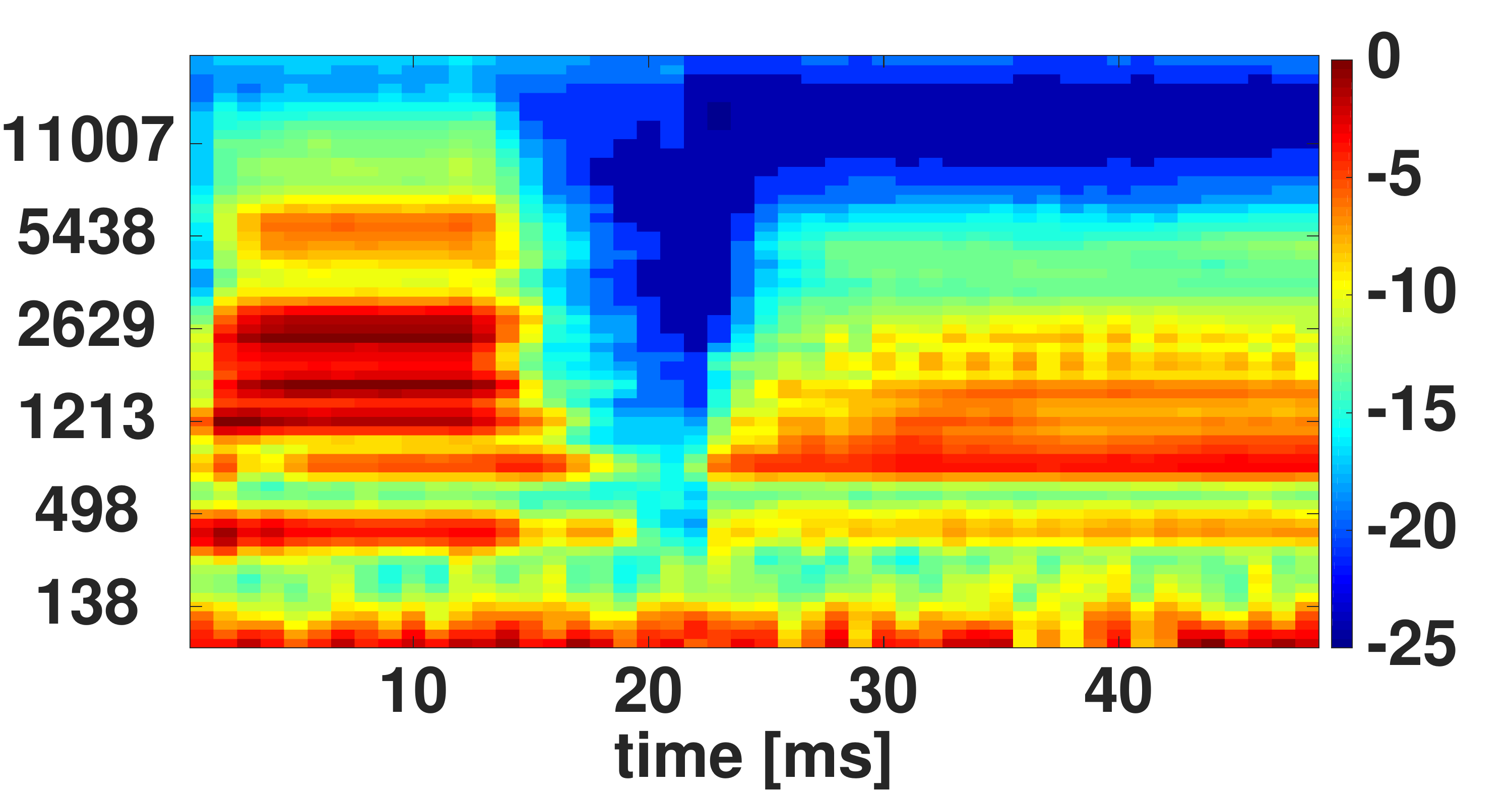}%
\caption{Horn}%
\label{subfig:gamma_others}%
\end{subfigure}%
\caption{Example of the gammatonegram representation of sound frames of $0.5s$  for the different acoustic classes considered, at different SNRs. From left to right: a yelp siren (SNR=-15dB), a wail siren (SNR=-10dB), and two different car horns (SNR=-13dB and SNR=-25dB, respectively). The energy of the time-frequency bins is expressed in dBFS. We observe that the frequency bins are not equally spaced, due to the application of the gammatone filterbanks.}
	\label{fig:gamma}
	\vspace{-2.5mm}
\end{figure*}

\subsection{Feature Representation}
\label{subsec:features}
Traditionally, audio classification relies on the use of Mel-Frequency Cepstrum Coefficients (MFCCs) \cite{zhuang2008feature}. Recent works \cite{marchegiani2017leveraging}\cite{chakrabarty2016abnormal},  demonstrated that MFCCs do not provide an acoustic signature which is robust to interference, leading to a deterioration in the classification performance, when operating in noisy scenarios. Given the potentially high level of noise we might encounter in traffic scenes, we here choose a different signal representation, based on the use of gammatone filterbanks \cite{lyon2010history}. Gammatone filterbanks have been originally introduced in \cite{holdsworth1988implementing}, as an approximation to the human cochlear frequency response, and, as such, can be used to generate a tempo-spectral representation of audio signals, where the frequency range is parsed in a \emph{human-like} fashion. The sounds we are interested in spotting are explicitly designed to be heard by humans, even in the presence of conspicuous traffic noise. Exploiting features mimicking human auditory perception, then, can be particularly convenient, as it is able to provide an additional pre-filtering of the signals.
The impulse response of a gammatone filter centred at frequency $f_c$ is:
\begin{equation}
    g(t, f_c)= 
\begin{cases}
    t^{a-1} e^{-2\pi bt} \cos{2\pi f_c t} & \text{if } t\geq 0\\
    0              & \text{otherwise}
\end{cases}
\end{equation}
where $a$ indicates the order of the filter, and $b$ is the bandwidth. The bandwidth increases as the centre frequency $f_c$ increases, generating narrower filters at low frequencies and broader filters at high frequencies. Following \cite{toshio1995optimal}, we utilise fourth-order filters (\textit{i.e.} $a=4$), and approximate $b$ as: 
\begin{equation}
b = 1.09 \; \left(  {f_c}/{9.26449}+24.7 \right) 
\label{eq:bandwidth}
\end{equation}
The centre frequencies are selected by applying the Equivalent Rectangular Bandwidth (ERB) scale  \cite{glasberg1990derivation}.  Let $x(t)$ be the original waveform audio signal, the output response $y(t, f_{c})$ of a filter characterised by the centre frequency $f_{c}$ can, then, be computed as:
\begin{equation}
y(t,f_c) = x(t) * g(t,f_c)
\label{eq:filter}
\end{equation}
Extending the filtering to the entire bank, across overlapping time frames, we obtain a \emph{gammatone}-like spectrogram, also known as \emph{gammatonegram}.    The gammatonegrams of a stereo combination (corresponding to two different receivers, \textit{i.e.} two different channels) of the original signals are computed and used in the semantic segmentation and event classification network, as well as in the acoustic source localisation network. Examples of gammatonegrams are provided in Fig. \ref{fig:gamma}.

\subsection{Acoustic Event Classification and Signal Denoising}
\label{subsec:mtl}
\begin{figure*}[t]
\vspace{2mm}
\centering
\begin{subfigure}{.45\columnwidth}
\includegraphics[width=\columnwidth]{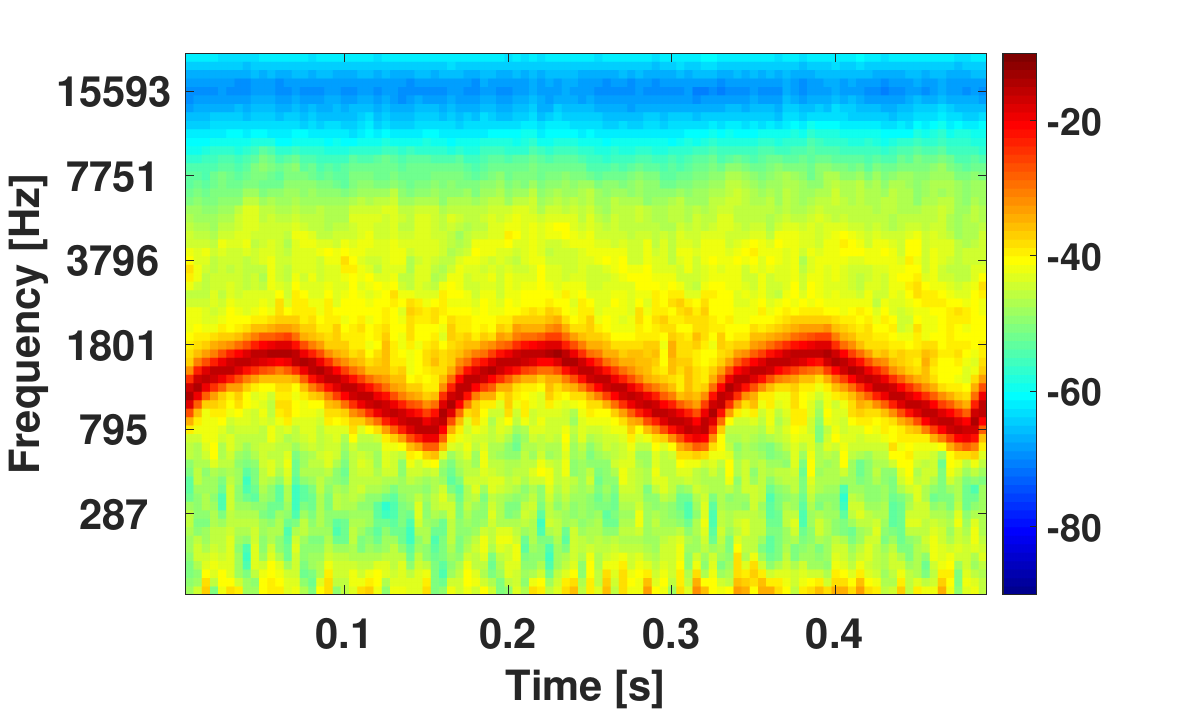}%
\caption{Original, SNR=5dB}%
\label{subfig:gamma_5}%
\end{subfigure}\hfill%
\begin{subfigure}{.45\columnwidth}
\includegraphics[width=\columnwidth]{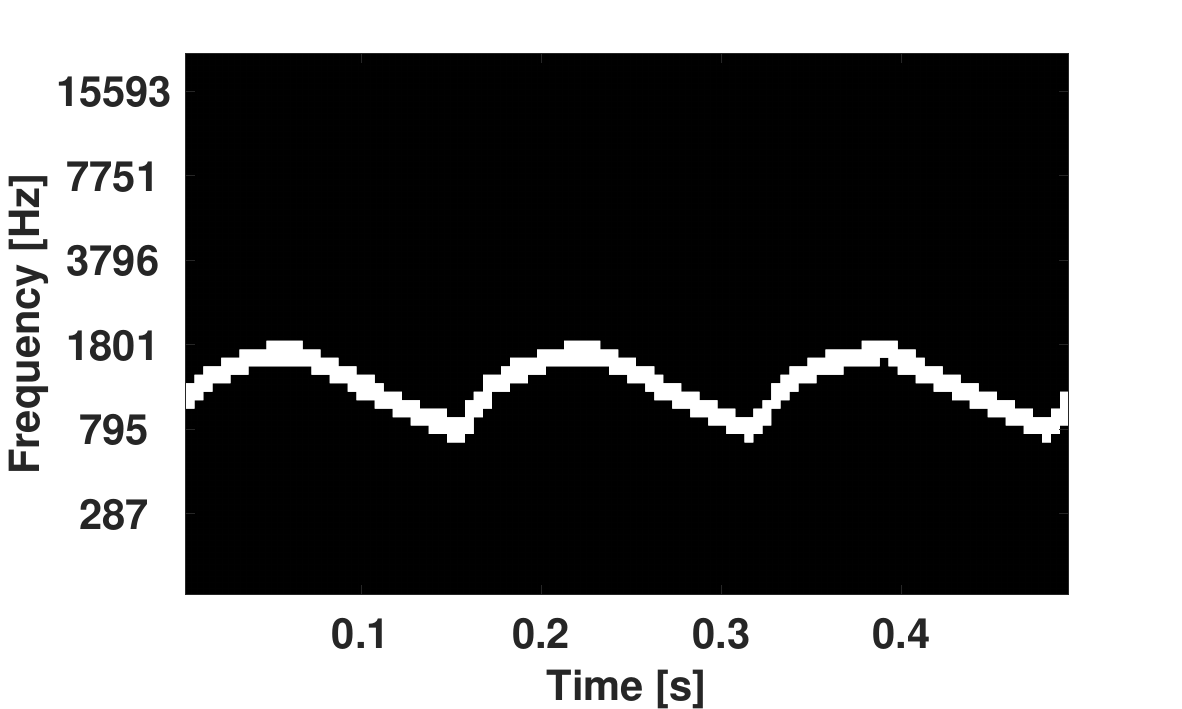}%
\caption{k-means, SNR=5dB}%
\label{subfig:emb_5}%
\end{subfigure}\hfill%
\begin{subfigure}{.45\columnwidth}
\includegraphics[width=\columnwidth]{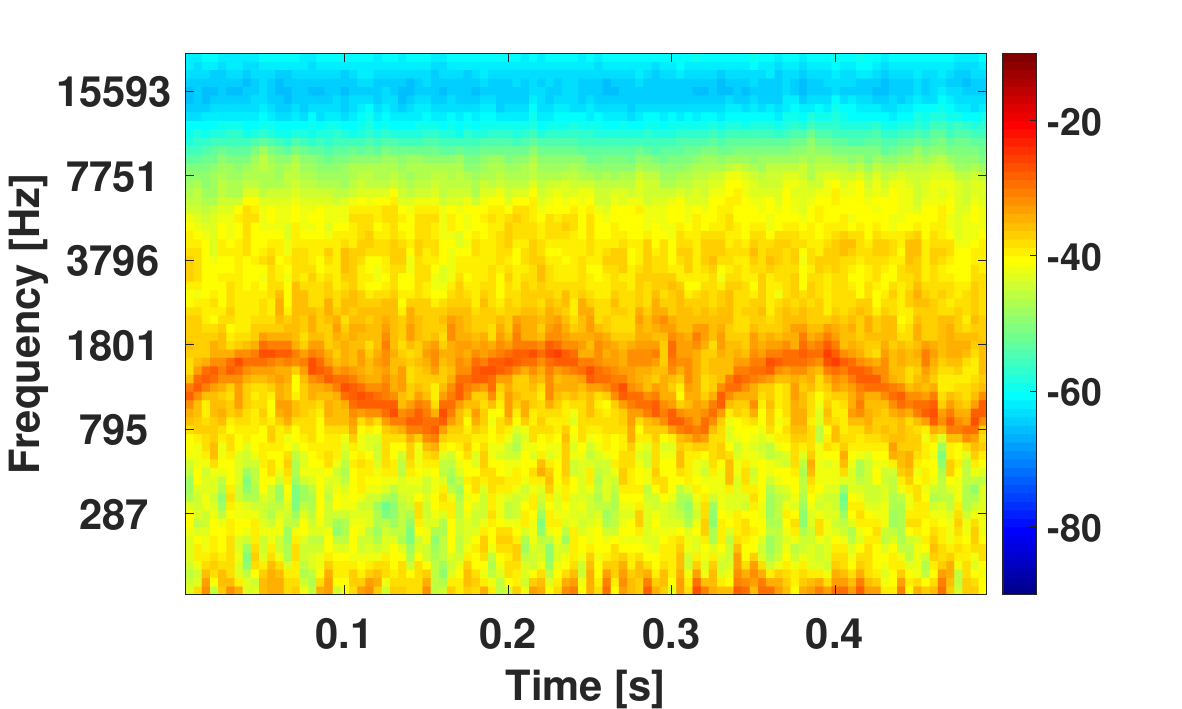}%
\caption{Original, SNR=-5dB}%
\label{subfig:gamma_m5}%
\end{subfigure}\hfill%
\begin{subfigure}{.45\columnwidth}
\includegraphics[width=\columnwidth]{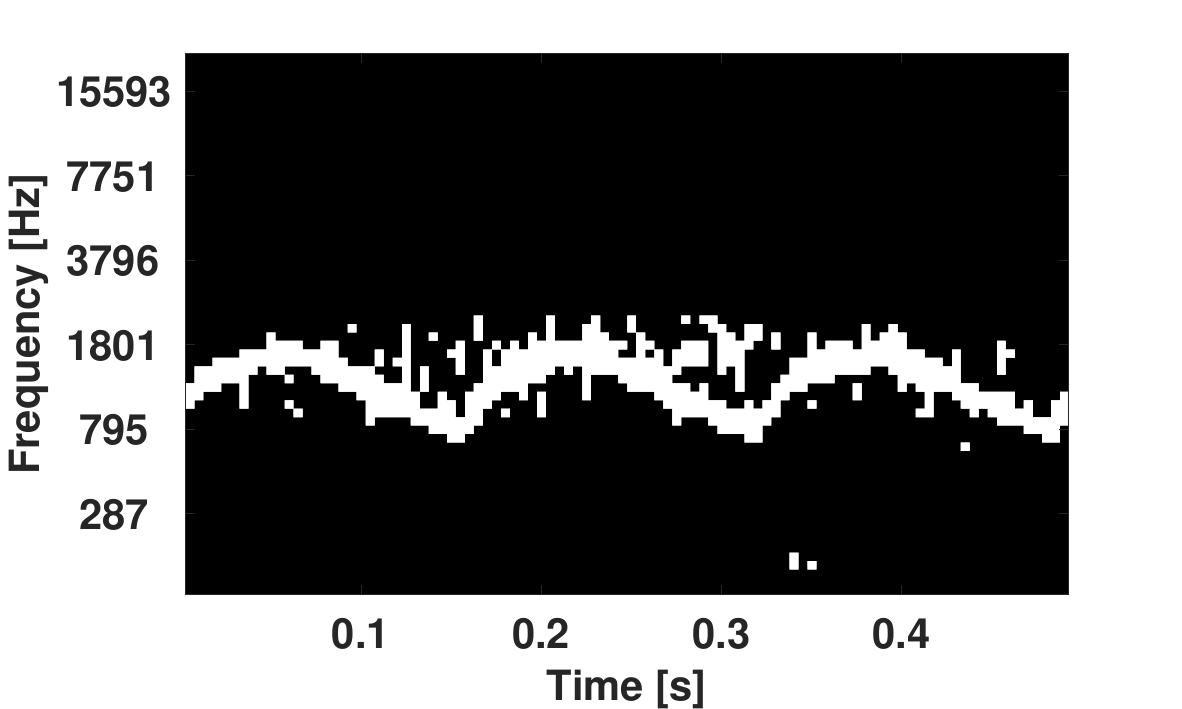}%
\caption{Segmentation, SNR=-5dB}%
\label{subfig:ebm_m5}%
\end{subfigure}%

\medskip
\begin{subfigure}{.45\columnwidth}
\includegraphics[width=\columnwidth]{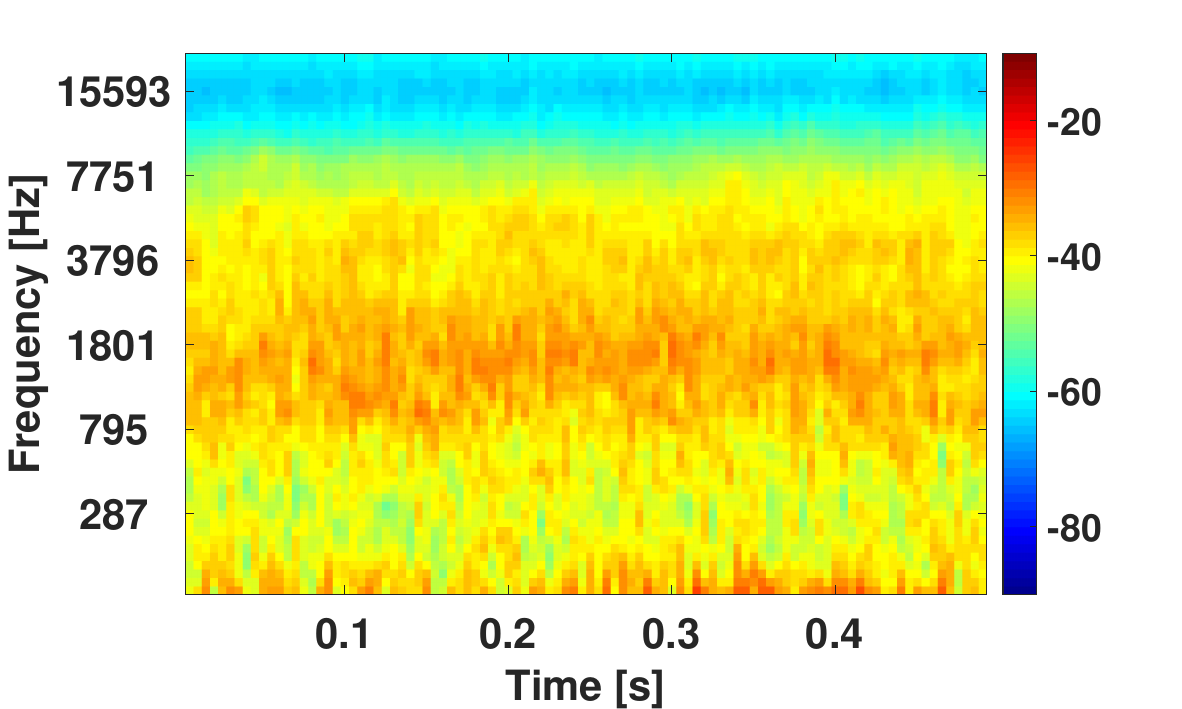}%
\caption{Original, SNR=-15dB}%
\label{subfig:gamma_m15}%
\end{subfigure}\hfill%
\begin{subfigure}{.45\columnwidth}
\includegraphics[width=\columnwidth]{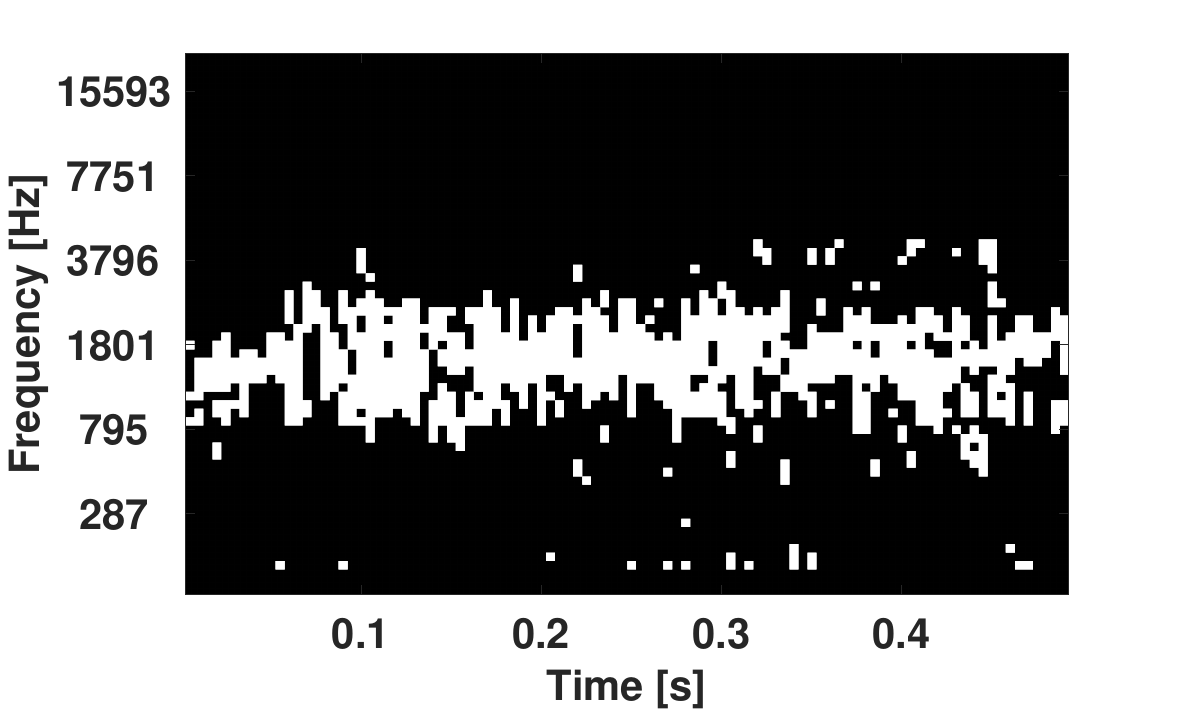}%
\caption{k-means, SNR=-15dB}%
\label{subfig:ebm_m15}%
\end{subfigure}\hfill%
\begin{subfigure}{.45\columnwidth}
\includegraphics[width=\columnwidth]{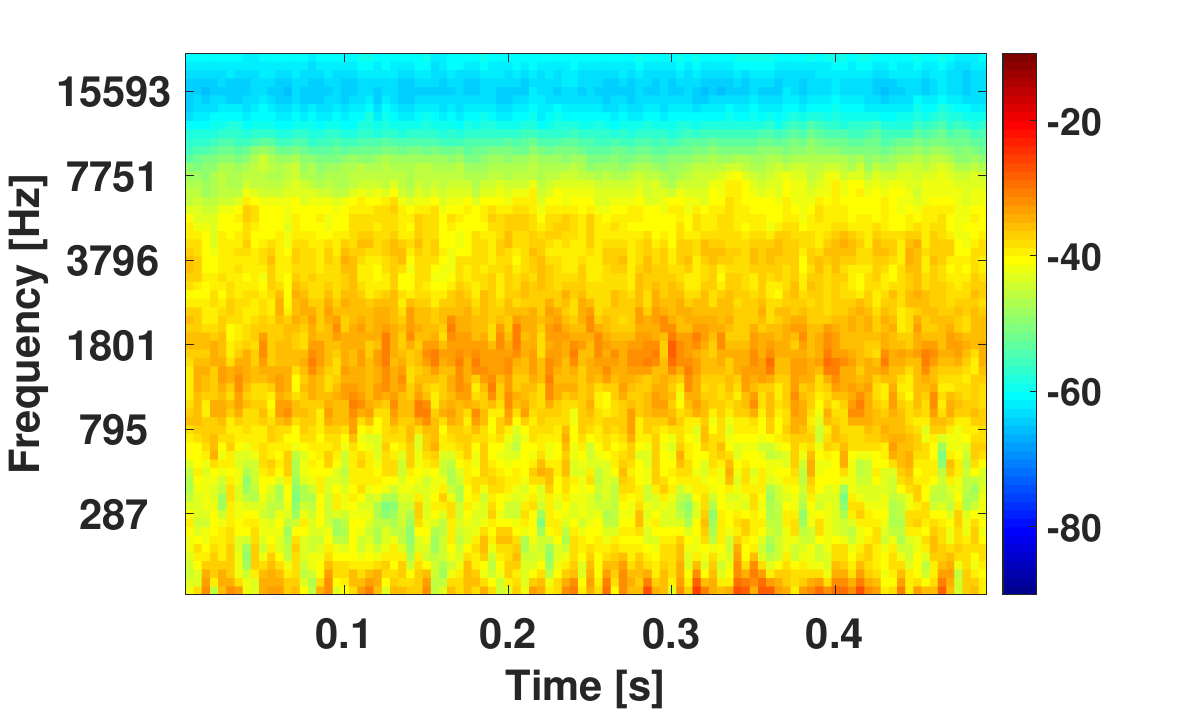}%
\caption{Original, SNR=-30dB}%
\label{subfig:gamma_m30}%
\end{subfigure}\hfill%
\begin{subfigure}{.45\columnwidth}
\includegraphics[width=\columnwidth]{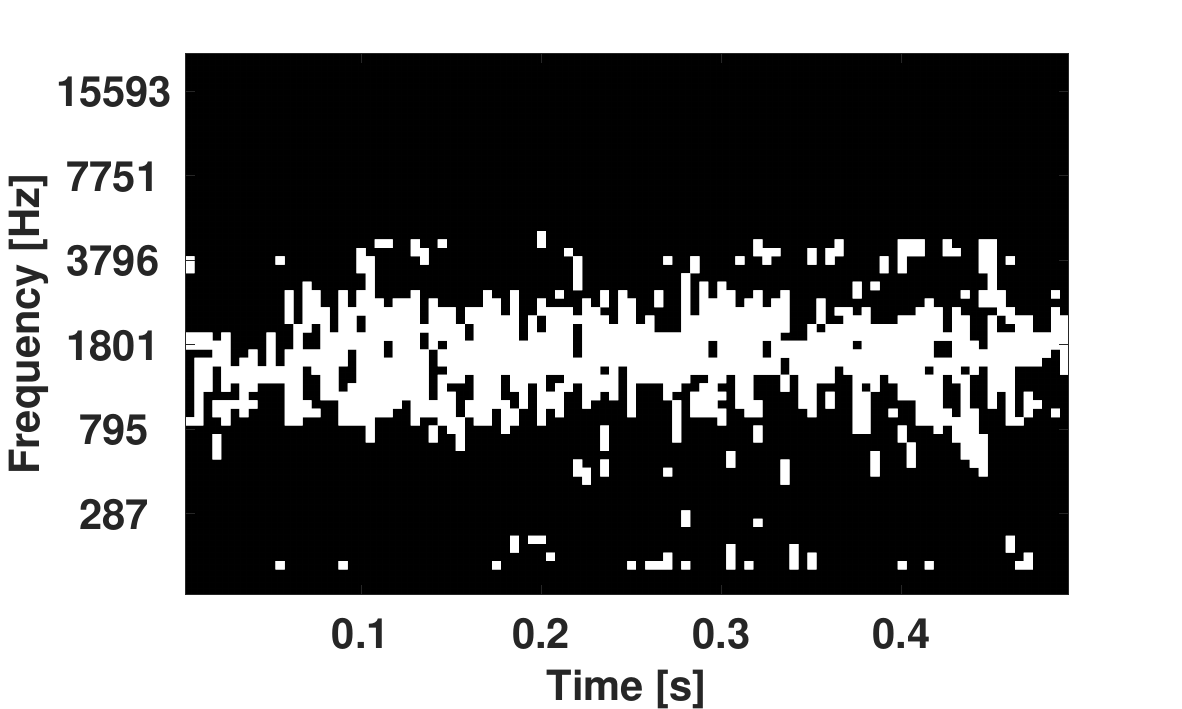}%
\caption{Segmentation, SNR=-30dB}%
\label{subfig:ebm_m30}%
\end{subfigure}%

\caption{Example \emph{k-means} segmentation, as described in \cite{marchegiani2017leveraging} operating on a gammatonegram representation of sound frames of $1 s$ corresponding to a yelp siren in background noise at different SNR levels. Original gammatonegram where $SNR=5dB$ (a) and corresponding segmentation mask (b); original gammatonegram where $SNR=-5dB$ (c) and corresponding segmentation mask (d); original gammatonegram where $SNR=-15dB$ (e) and corresponding segmentation mask (f); original gammatonegram where $SNR=-30dB$ (g) and corresponding segmentation mask (h).}
	\label{fig:kmeans}
	\vspace{-2.5mm}
\end{figure*}

As mentioned in Section \ref{sec:intro}, in our previous work \cite{marchegiani2017leveraging}, we perform gammatonegram segmentation employing \emph{k-means}. \emph{k-means} has the advantage to operate in a completely unsupervised manner (\textit{i.e.} no training or labelling necessary). Nevertheless, some parameters still need to be predefined, such as the number of clusters (\textit{i.e.} parts we want the image to be segmented into), which is not a trivial choice to make in this particular scenario, where the noise is of a variegate structure and nature. Furthermore, it also necessary to make assumptions  to decide which of the clusters actually corresponds to the target signal. In our previous framework, we assume the cluster corresponding to the most powerful time-frequency slots in the gammatonegrams to be the target signal. Yet, this assumption cannot hold in really low SNRs, where the most powerful time-frequency slots in the gammatonegrams might actually belong to the background noise.  One might argue that cluster assignment could be based on the frequency range where the target signal lies. However, while this could be, perhaps, applied in the case of sirens, where the frequency content is clearly defined, the same can not be said about horns, for instance. Additionally, in our previous system, the segmentation phase precedes the classification one. This entails that, when assigning clusters, we would not be able to look for the frequency content of a specific signal, as we would not know yet which signal to target. All of the shortcomings, described above, greatly worsen when dealing with scenarios characterised by really low SNRs. Fig.~\ref{fig:kmeans} shows qualitative examples of the change in the performance of the \emph{k-means} approach, at various levels of background noise.

In this work, we perform acoustic event classification and signal denoising utilising a multi-task learning (MTL) scheme. Multi-task learning has been successfully employed following various implementation strategies, and in several domain applications, such as language processing \cite{collobert2008unified} and traffic flow forecasting \cite{jin2008neural}. By taking advantage of information in training samples of related tasks, MTL has proved to be a valuable tool for reducing overfitting and, consequently, improving the models' generalisation capabilities. In this work, we opt for \emph{hard parameter sharing}, which was \rrv{first} introduced in \cite{caruana1998multitask}, having the tasks directly share some of the architecture. 

We implement noise removal, by treating the gammatonegrams of the incoming sound as intensity images, and feeding them to a \emph{Unet} \cite{long2015fully}, to carry out semantic segmentation. Specifically, we rely on an architecture similar to the one defined in \cite{ronneberger2015u}. As the result of the segmentation will later be used to localise the sound source, we use as input to the network a concatenation of the gammatonegrams of the two signals in the stereo combination, rather than analysing one channel at a time. We make this choice, as we hypothesise it will help capture inter-channel information, and allow us to obtain a more \emph{stereo-aware} segmentation. The \emph{Unet} can be seen as an autoencoder, relying on two main processing phases: encoding, and following decoding. The encoding generates a more compact representation of the input (\textit{i.e.}  the code), while the decoding attempts to reconstruct the original input, filtered depending on the specific task. In this case, the output of the decoding step will be a segmented version of the original gammatonegrams. In order to perform acoustic event classification, the network is augmented by fully connected layers, which, operating on the \emph{code} generated in the encoding step, assign the input signal to one of three classes of interest: siren of an emergency vehicle, horn, and any other kind of traffic noise. The MTL scheme aims to simultaneously assign a label (\textit{i.e.} siren, horn, other sound) to each pixel (corresponding to a time-frequency slot of the gamamtonegrams) of the input image (\textit{i.e.} segmentation), and to the entire image (\textit{i.e.} event classification). 

The complete structure of the network is reported in Fig.~\ref{fig:architecture}. 
The encoding part of the network (\textit{i.e.} left side of the network) consists of three layers, where each layer presents the application of two successive $3 \times 3$ convolutions, followed by a $2 \times 2$ max pooling operation, with stride 2 for downsampling. The first layer is characterised by $64$ feature channels, which double at each layer. The decoding part of the network (\textit{i.e.} right side of the network), instead, presents a $2 \times 2$ upsampling step, which reduces in half the number of feature channels, followed by the application of two successive  $3 \times 3$ convolutions. After each upsampling operation, the resulting feature map is concatenated to the respective one from the left side of the network. All convolutions occur with an Exponential Linear Unit (ELU) \cite{ClevertUH15}. A final $1 \times 1$ convolution is applied to assign a label to each pixel. Acoustic event classification is obtained by adding two fully connected layers at the bottom of the \emph{Unet} (\textit{cf.} yellow area in Fig.~\ref{fig:architecture}).  
We train the multi-task learning architecture by minimising the loss $L_{SeC}$, defined as:
\begin{equation}
    L_{SeC} = L_S + L_C  
    \label{eq:cost_mtl}
\end{equation}
where $L_S$ and $L_C$ refer to the loss related to the segmentation and classification tasks, respectively, and are computed by applying a soft-max combined with a cross-entropy loss function. In the case of $L_S$, the soft-max is applied pixel-wise over the final feature map, and  defined as:
\begin{equation}
p_{i,k} = \frac{\exp(a_{i,k})}{\sum_{n=1}^{N}\exp(a_{i,k})}
\end{equation}
where $a_{i,k}$ indicates the activation in feature map $k$ at pixel $i$, and $N$ is the number of classes. Training is performed by minimising the loss $L_{SeC}$ with $l1$ regularisation, using back-propagation.

\begin{figure*}
\centering
\begin{subfigure}{.33\textwidth}
\centering
\includegraphics[width=0.7\columnwidth]{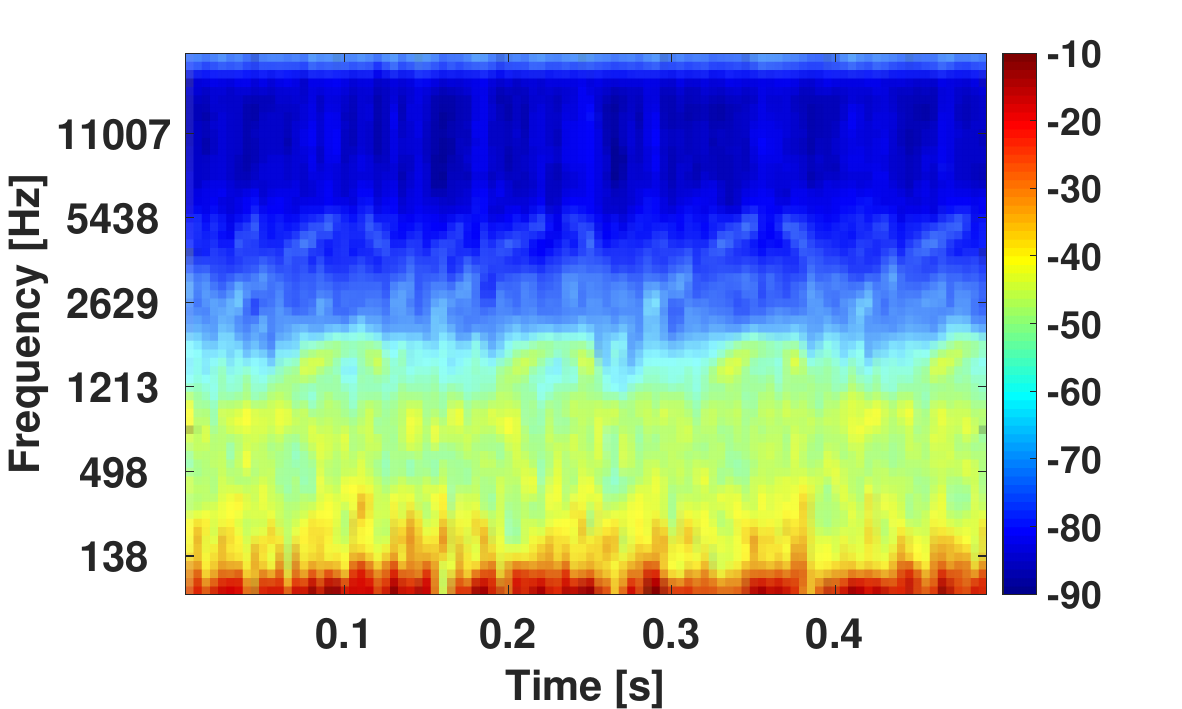}%
\caption{Original, SNR=-35dB}%
\label{subfig:gamma_original}%
\end{subfigure}\hfill%
\begin{subfigure}{.33\textwidth}
\centering
\includegraphics[width=0.7\columnwidth]{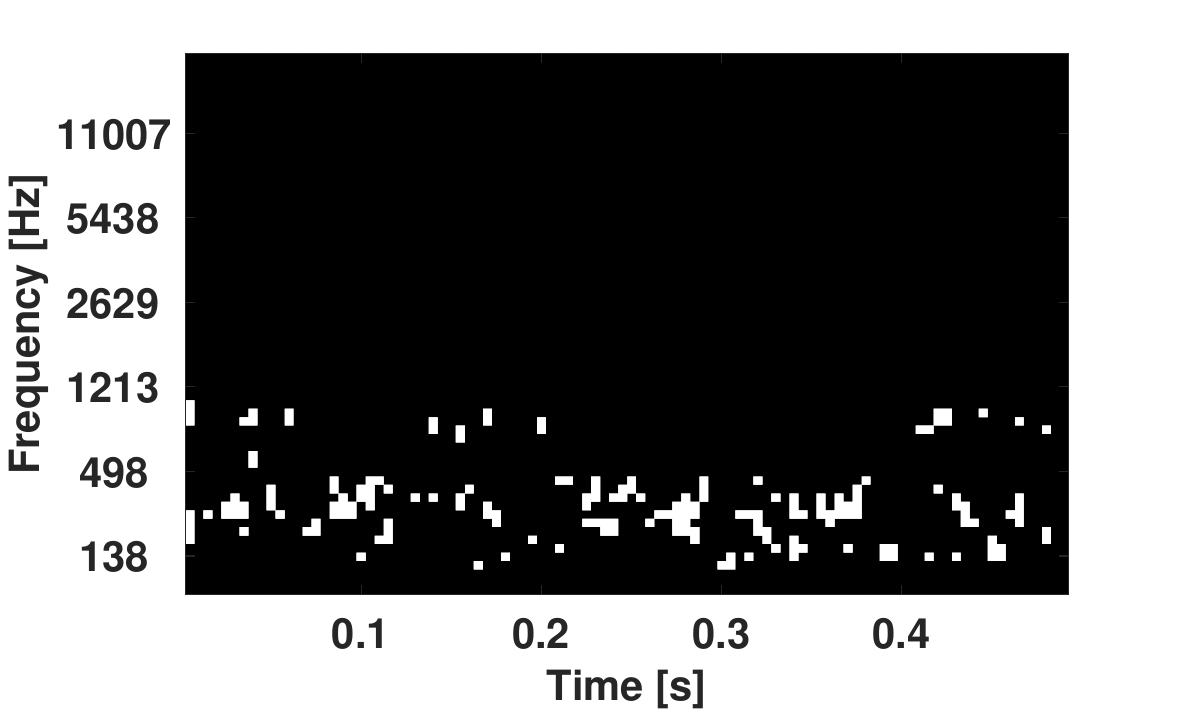}%
\caption{k-means, SNR=-35dB}%
\label{subfig:kmeans}%
\end{subfigure}\hfill%
\begin{subfigure}{.33\textwidth}
\centering
\includegraphics[width=0.7\columnwidth]{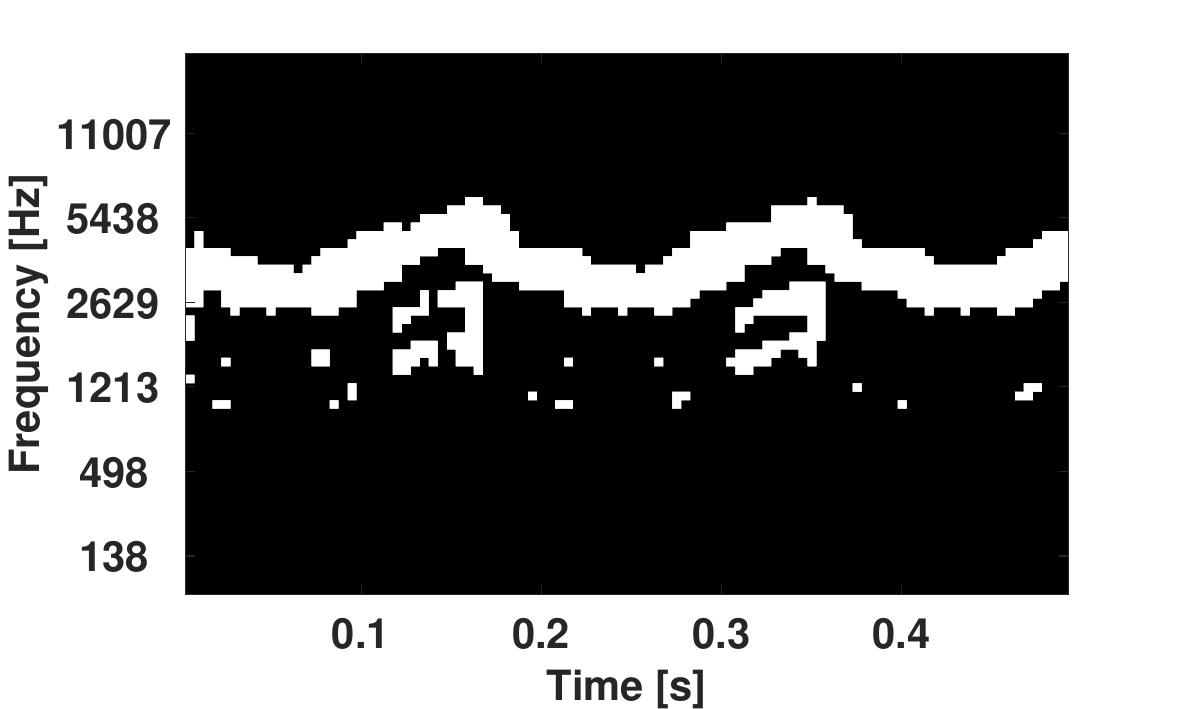}%
\caption{Unet, SNR=-35dB}%
\label{subfig:unet_seg}%
\end{subfigure}
\caption{The figure shows a qualitative comparison between the \emph{k-means} approach and the \emph{Unet} one, when operating on a gammatonegrams characterised by an $SNR$ level of $-35dB$. From left to right: the original noisy gammatonegram (a), the segmentation map provided by the \emph{k-means} method\cite{marchegiani2017leveraging} (b), and the one obtained by applying the \emph{Unet} approach.}
	\label{fig:kmeans_comparison}
	\vspace{-2.5mm}
\end{figure*}

One of the reasons for which the \emph{k-means} approach results to be less resilient to greater amount of background noise is that, in that case, the segmentation is merely based on the energy content of each time-frequency slot of the gammatonegram, while, when applying a \emph{Unet}, as we operate in a supervised manner, we are actually able to learn, look for, and retrieve the particular shape of the signal of interest. As we are now operating in a MTL scheme, the segmentation is also augmented by information given by the classification task, lessening some of the issues induced by performing a \textit{blind} segmentation (\textit{i.e.} we first segment the spectrogram and then classify the signal; the segmentation can not make use of any signal-specific information), as mentioned earlier. Fig.~\ref{fig:kmeans_comparison}
reports a qualitative comparison between the \emph{k-means} approach and the \emph{Unet} one, when applied to a scenario characterised by an SNR of $-35$dB. We can observe that even if the deep learning-based system does not recover perfectly the original target signal, the segmentation is still much more accurate and reliable than the one provided by \emph{k-means}\cite{marchegiani2017leveraging}.

\subsection{Sound Source Localisation}
\label{subsec:ssl}
In this work, we are interested in horizontal acoustic source localisation, relying on a stereo composition of the sound (\textit{i.e.} as perceived by two different, spatially separated, microphones). Specifically, we are interested in learning a direct mapping between the \emph{clean} gammatonegrams of the stereo signal and the direction of arrival of the sound. Once the segmentation has been performed, as described in Section~\ref{subsec:mtl}, the gammatonegrams of two channels of the target signal are recovered by applying the output of the segmentation as a mask on the original gammatonegrams. The cross-correlation between the resulting clean gammatonegrams, the \emph{cross-gammatonegram} is then used as input to a CNN to regress the DoA. If $G_{N_1}$  and $G_{N_2}$ are the noisy gammatonegrams for the first and second channel of the stereo signal, and $S_1$ and $S_2$ the respective segmented images, the input to the network is given by:
\begin{equation}
\begin{split}
( G_{C_1} \star G_{C_2} ) (p,l)= \\
\sum_{m=0}^{M-1} \sum_{n=0}^{N-1} G_{C_1}(m,n) \cdot \bar{G}_{C_2}(m-p,n-l), \\ \
-M+1 \leq p \leq M-1, \\
-N+1 \leq l \leq N-1
\end{split}
\end{equation}
where $G_{C_j} = G_{N_j} \circ S_j  / \max{ \left (G_{N_j} \circ S_j \right ) }, \ j \in \{1,2\}$
and $\bar{G}_{C_j}$ denotes the complex conjugate. 
The CNN consists of two $6 \times 6$ convolutional layers, followed by a  $2 \times 2$ max pool, and two fully connected layers.  All layers are equipped with an ELU. 
We employ the network to regress the direction of arrival of the sound as the respective angle on the horizon plane, and define our loss function $L_{SL}$ as
$    L_{SL} = \| \alpha - \hat{\alpha} \|_2 $,
where $\alpha$ and $\hat{\alpha}$ are the ground truth values, and the predictions of the network.  Training is performed by minimising the Euclidean loss $L_{SL}$ with $l1$ regularisation, using back-propagation.

\section{EXPERIMENTAL EVALUATION}
\label{sec:exp}

We evaluate our framework, analysing the performance of both networks, comparing their behaviour with other two different architectures. We show that the particular configuration chosen in this work, while providing comparable performance in the acoustic event classification task, yields significantly greater performance in the sound source localisation one compared to the other architectures analysed. Specifically, we evaluate two alternative networks:
\begin{itemize}
\item \emph{Full-Sharing (FS)}: the two tasks in the MTL scheme share both the enconding and decoding side of the network. Classification does not take place at the bottom of the \emph{Unet}, as in Fig. \ref{fig:architecture}, but at the last decoding layer.
\item \emph{Mono}: the multi-task learning scheme is identical to the one of the \emph{SeC} network, but, in this case, it does not operate on the gammatonegrams of the stereo sound, but gammatonegrams of different channels are considered as separate samples.
\end{itemize}
We remind the reader that we apply semantic segmentation to the gammatonegrams of the stereo sound as a denoising technique, which allows us to recover the target signal from the background noise. As such, we are not interested in the performance of the segmentation \textit{per se}, but rather in the accuracy we obtain when regressing the DoA from the gammatonegrams which have been \emph{cleaned} by the segmentation. Thus, in this context, we will focus on the performance of the \emph{SL} network and the classification task of the \emph{SeC} one, and discuss the performance of the semantic segmentation only through the impact this has on the following DoA estimation.

\vspace{-0.3cm}
\subsection{The Dataset}
\label{sub:exp:dataset}
\begin{figure}
\centering
\includegraphics[width=0.6\columnwidth]{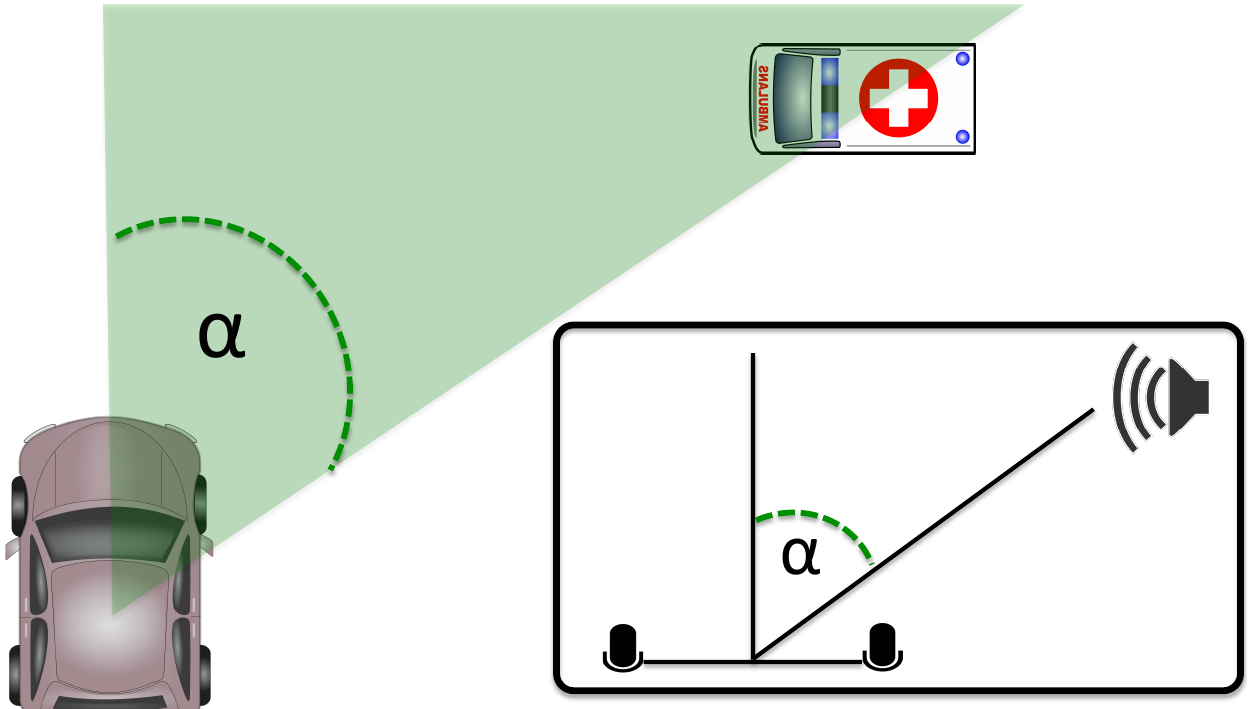}%
\caption{Microphone configuration used to generate the dataset. The two microphones are mounted on the roof of the vehicle. The DoA of the sound, computed as the angle between the sound source and the vehicle, is denoted by $\alpha$.}
\vspace{-4mm}
\label{fig:mics}
\end{figure}

To evaluate the performance of our framework, we collected four hours of data by driving around Oxford, UK, on different kinds of road, and at different times of the day (\textit{i.e.} different traffic conditions). The data was gathered using two Knowles omnidirectional boom microphones mounted on the roof of the car and an ALESIS IO4 audio interface. The data was recorded at a sampling frequency $f_s$ of $44100$ Hz  at a resolution of $16$ bits. Furthermore, to obtain accurate ground truth values in the sound source localisation task against various levels of masking noise, we corrupted a stereo composition of specific target signals with the traffic noise recorded, generating samples at various SNRs. This kind of approach is commonly used in acoustics literature (\textit{e.g.} \cite{chan2016listen}\cite{ marchegiani2015cross}\cite{schroder2013automatic}\cite{noda2015sound} among others), especially when the impact of noise on classification and identification tasks has to be isolated and accurately quantified. Lastly, it allows us to address scenarios where no other sensors can be used to provide ground truth, as either the target sound is purely acoustic (\textit{e.g.} horns), or the sound source is too far away and, thus, out of the field of view of additional sensors potentially present (situation which would, indeed, match the low SNR levels analysed in this work). This additional data used has been extracted from the Urban Sound Dataset \cite{Salamon:UrbanSound:ACMMM:14}, and from other publicly available databases, such as \texttt{\url{www.freesound.org}}. We are interested in clean signals, as these will represent our ground truth data. Thus, we select only samples, where any background noise is either absent or can be easily removed through traditional filtering. The clean signals obtained are then mixed with the recorded traffic noise, simulating different direction of arrival of the sound, with the acoustic source moving at different velocities, following different paths. Frequency shifts, due to the Doppler effect, are applied accordingly. The simulation also takes into account additional propagation effects, such as echoes (\textit{i.e.} delayed, less powerful copies of the original signal), and small perturbations (\textit{i.e.} variation in the power of the perceived signal depending on the direction of arrival) to consider potential reflections, and different kinds of microphone response patterns. A schematic representation of the microphone configuration is given in Fig. \ref{fig:mics}. The direction of arrival of the sound, computed as the angle between the sound source and the vehicle is denoted by $\alpha$. \rv{The data were collected with the microphones positioned at a fixed location, due to physical constraints associated with mounting microphones on a moving car safely}. The current framework operates on a $180 \degree$ space; yet it can be easily extended to $360 \degree$ with an additional microphone.
In total we generate more than $30K$ samples of $0.5s$ frames, equally distributed among the three classes: sirens, horns, and others (\textit{i.e.} any other traffic sound, which is neither a siren nor a horn).

\begin{figure*}[ht!]
  \begin{tabularx}{\linewidth}[t]{*{3}X}
\resizebox{0.55\columnwidth}{!}{%
\renewcommand{\arraystretch}{1}
\begin{tabular}{c c c c }
\toprule 
& \multicolumn{3}{c}{Predicted Class} \\
\cline{2-4}
\multicolumn{1}{c }{True Class} & \multicolumn{1}{c}{Siren} & \multicolumn{1}{c}{Horn} & \multicolumn{1}{c}{Other} \\ \midrule
Siren  & \textbf{0.98}  & 0.02   & 0  \\ 
Horn   & 0.10   & \textbf{0.90}   & 0 \\ 
Other  & 0  & 0.06   & \textbf{0.94}  \\ \bottomrule
\end{tabular}
}
&
    \centering
    \begin{tabular}[c]{c}
      \includegraphics[width=0.6\columnwidth]{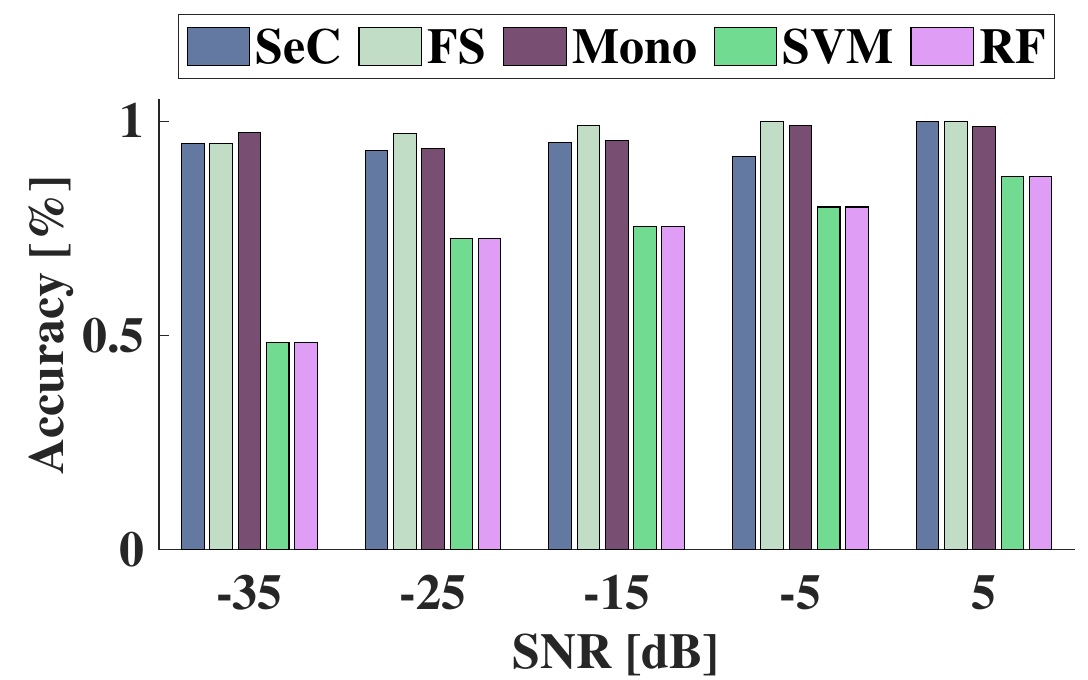}
    \end{tabular} &
      \centering
\resizebox{0.5\columnwidth}{!}{%
\renewcommand{\arraystretch}{1}
\begin{tabular}{c c c c}
\toprule 
&\multicolumn{3}{c}{Regression Error} \\
\midrule
\multicolumn{1}{c }{\multirow{2}{*}{}}& $\tilde{E}_{SeC}$ &$\tilde{E}_{FS}$ & $\tilde{E}_{Mono}$\\ \cline{2-4}
Siren  & \textbf{6.4}& 10.4 & 7.9  \\ 
Horn  & \textbf{8.8} & 18.9 & 20.9  \\
All & \textbf{7.5} & 14.1 & 12.9 \\\bottomrule
\end{tabular}
}
    \tabularnewline
    \captionof{table}{Confusion Matrix obtained by employing the \emph{SeC} network. The average classification rate for all classes is shown along the diagonal of the matrix.} \label{table:conf_mat} &
    \captionof{figure}{\rrv{Classification accuracy, averaged over the three classes, at various SNR levels, when applying the \emph{SeC}, the \emph{FS}, the \emph{Mono} networks, and the \emph{SVM} and \emph{RF}-based  systems proposed in \cite{Salamon:UrbanSound:ACMMM:14}.}}       \label{fig:class_snr}
    &
    \captionof{table}{Median absolute error in the DOA, when performing the segmentation with the \emph{SeC}  ($\tilde{E}_{SeC}$), \emph{FS} ($\tilde{E}_{FS}$), and the \emph{Mono} ($\tilde{E}_{Mono}$) networks.}
    \label{table:exp_reg}\tabularnewline
\end{tabularx}
\vspace{-5.5mm}
\end{figure*}

\vspace{-0.3cm}
\subsection{Implementation Details}
We trained the networks using mini-batch gradient descent based on back propagation, employing the Adam optimisation algorithm \cite{kingma2014adam}. We applied dropout \cite{hinton2012improving} to each non-shared layer for both tasks' architectures with a keeping probability of $0.9$. The models were implemented using the Tensorflow \cite{tensorflow2015-whitepaper} libraries. We confine our frequency analysis to a range between $50$ Hz and $f_s/2 = 22050$ Hz, corresponding to the maximum reliable frequency resolution available, and utilise $64$ frequency channels in the gammatone filterbank. The filtering is computed on time domain frames of $0.5s$ with $10 ms$  overlap, after applying a Hamming window to avoid spectral leakage. Similarly to previous works on deep learning in the auditory domain (cf. \cite{deng2014robust}, \cite{takahashi2017aenet}), we randomly split our dataset into training set ($90\%$) and test set ($10\%$).

\vspace{-0.3cm}
\subsection{Acoustic Event Classification}
\label{subsec:sec}
Table \ref{table:conf_mat} reports the confusion matrix obtained by employing the  \emph{SeC} network. The average classification rate for all classes is shown along the diagonal of the matrix.  Fig. \ref{fig:class_snr} shows the accuracy obtained in the classification, averaged over the three classes, at various SNR levels, when applying the \emph{SeC} network, and the two benchmarks: the \emph{FS}, and the \emph{Mono} networks. \rrv{(The figure also includes results that will be discussed in Section~\ref{subsec:discussion}.)} The results suggest that all three architectures are able to provide a great classification accuracy, despite the presence of copious noise in the original gammatonegrams. Specifically, \emph{SeC} provides an average accuracy of $94\%$, while \emph{FS} and \emph{Mono} provide an accuracy of $98 \%$ and $96\%$, respectively. Furthermore, Table \ref{table:conf_mat} confirms that this accuracy is stable among the different classes.

\vspace{-0.3cm}
\subsection{Sound Source Localisation}
\label{subsec:ssl}
\begin{figure*}[ht!]
  \begin{tabularx}{\linewidth}[t]{*{3}X}
  \centering
    \begin{tabular}[c]{c}
      \includegraphics[width=0.6\columnwidth]{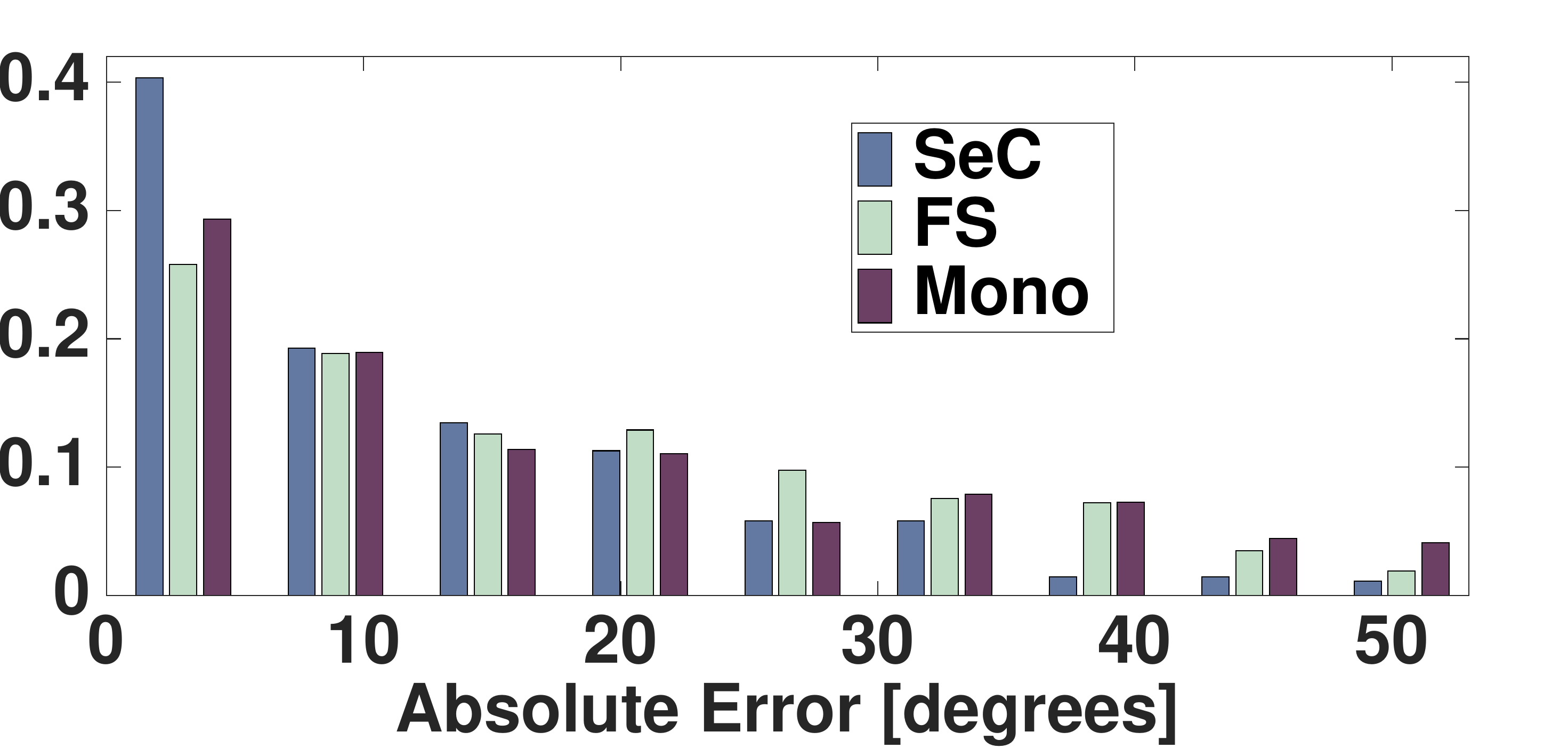}
    \end{tabular}
&
    \begin{tabular}[c]{c}
      \includegraphics[width=0.55\columnwidth]{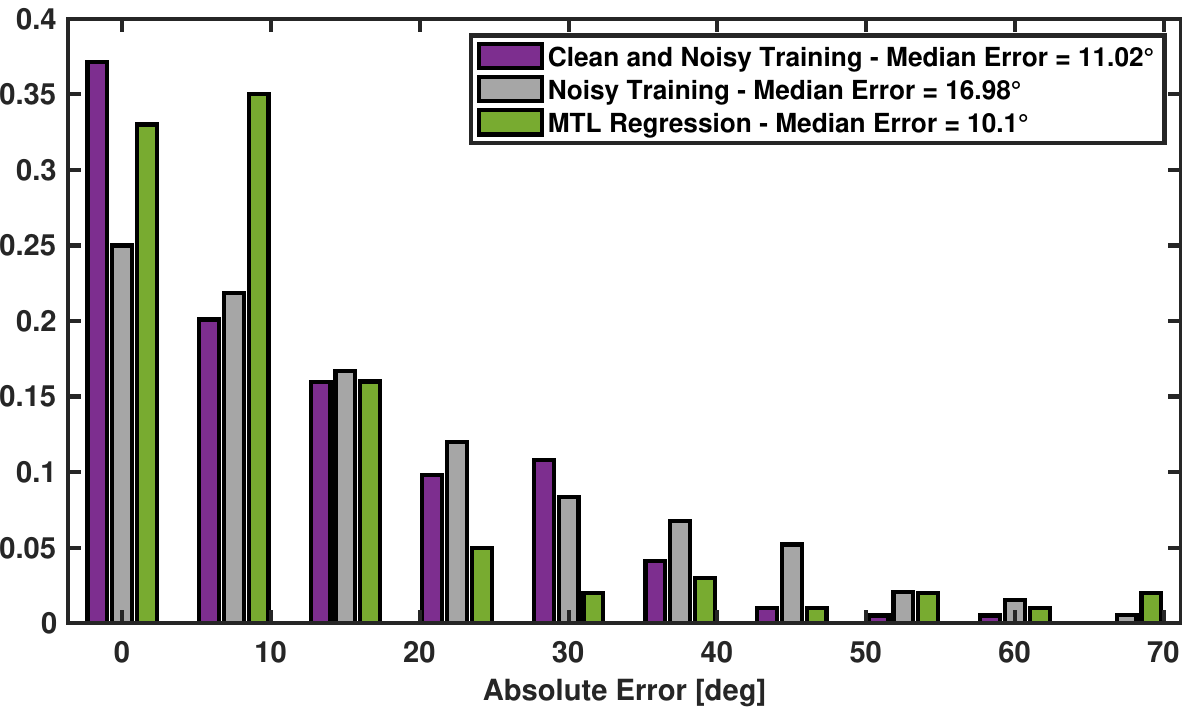}
    \end{tabular} &
    \centering
    \begin{tabular}[c]{c}
      \includegraphics[width=0.66\columnwidth]{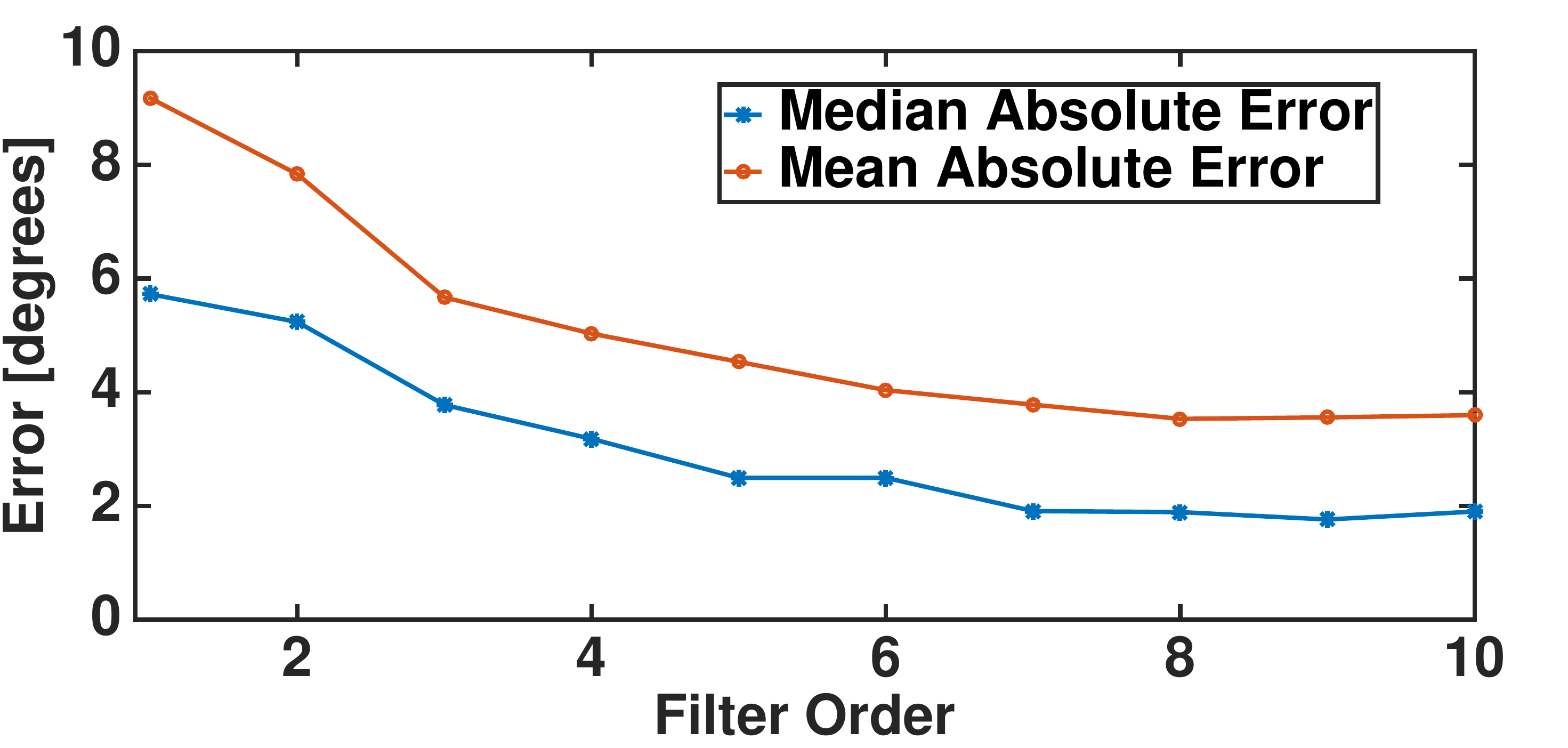}%
    \end{tabular} 
    \tabularnewline
    \captionof{figure}{Normalised histograms of the absolute error in the estimation of the DoA, with segmentation through the \emph{SeC}, \emph{FS}, and \emph{Mono} networks.}
\label{fig:reg_hist} &
    \captionof{figure}{\rv{Normalised histograms of the absolute error in the DoA, when using the two \emph{end-to-end} CNNs on noisy signals, and  \emph{MTL Regression}.}}
\label{fig:reg_hist_no_segmentation}
    &
    \captionof{figure}{Mean and median absolute error obtained by applying a median filter of a different order to the DoA estimates.}
\label{fig:reg_med}\tabularnewline
\end{tabularx}
\vspace{-0.5cm}
\end{figure*}

Table \ref{table:exp_reg} reports the median absolute error obtained in the regression of the DoA, when the preliminary segmentation is carried out through the \emph{SeC} network (\textit{cf.} $\tilde{E}_{SeC}$), and through the two benchmarks: the \emph{FS} (\textit{cf.} $\tilde{E}_{FS}$), and the \emph{Mono} networks (\textit{cf.} $\tilde{E}_{Mono}$). The correspondent normalised histograms of the absolute error are shown in Fig. \ref{fig:reg_hist}.  In all three cases, the DoA estimation relies on the \emph{SL} network. We observe that our framework is able to accurately localise the acoustic source, successfully coping with scenes characterised by extremely low SNRs ($-40 dB \leq SNR \leq 10  dB$). Furthermore, it is the one yielding the greatest performance, when compared to the \emph{FS} and the \emph{Mono} networks. In particular, while losing only $4\%$, and $2\%$ in the event classification task accuracy, we obtain a $46\%$ improvement on $\tilde{E}_{FS}$ and $42 \%$  on $\tilde{E}_{Mono}$ in the DoA estimation, when considering both sirens and horns. We conclude that:
\begin{itemize}
\item \emph{SeC vs FS}: performing classification at the bottom of the \emph{Unet} allows the segmentation to learn more task-specific patterns, yielding an increase in the DoA estimation accuracy. 
    \item  \emph{SeC vs Mono:} performing segmentation on the stereo gammatonegrams, rather than on the one of each channel independently, allows us to learn a more \emph{stereo-aware} representation of the sound, which makes the DoA regression more robust and accurate.
\end{itemize}
We also observe that, in all the frameworks, the estimation process is more accurate with the sirens, than with the horns. This is as expected, as horns have characteristics similar to the ones of pure tones, and consist of a dominant frequency component, whose patterns tend to variate, only slightly, over time.   Such characteristics reduce the ability of the system to detect the auditory cues necessary to correctly regress the direction of arrival of the sound, which are based on the difference between the gammatonegrams of the two signals in the stereo combination. From Table \ref{table:exp_reg}, we see that, in the case of the horns, employing \emph{SeC} for segmentation provides a $53\%$ improvement on $\tilde{E}_{FS}$ and $57 \%$  on $\tilde{E}_{Mono}$.

In our previous work \cite{marchegiani2017leveraging} we demonstrated that spectrogram segmentation greatly improves the accuracy in the classification of the various acoustic events. Similarly, we here investigate the impact of the segmentation on sound source localisation. Specifically, we evaluate the behaviour of two additional CNN architectures, which implement \emph{end-to-end} regression, \textit{i.e.} from the generalised cross-correlation of the noisy original gammatonegrams of the signals to the direction of arrival of the sound. The structure of the CNN is the same as for the \emph{SL} network, where the input is given by the generalised cross-correlation of the noisy gammatonegrams of the signals without applying any pre-processing and segmentation. The only difference in the two architectures is in the datasets used for training. The first network is trained using a dataset of noisy and clean gammatonegrams, while the second one is trained using a dataset of only noisy gammatonegrams. We tried both approaches to also verify whether by exposing the network to  clean gammatonagrams could help it learn how to implicitly discard the noise. Testing is always performed on noisy signals. \rv{Furthermore, we analyse the behaviour of an additional framework where no segmentation is performed, but event classification and sound source localisation are carried out in a multi-task learning fashion, which we call \emph{MTL Regression}. In  particular, this MTL scheme combines the \emph{SL} network where neither pre-processing or segmentation are applied and a CNN structure similar to the one used for class prediction in the \emph{SeC} architecture. The goal was to investigate whether supplying the network with information about the class of the signal could ease the regression task.}
Fig.~\ref{fig:reg_hist_no_segmentation} reports the normalised histograms of the absolute error obtained \rv{in all three scenarios}, together with the corresponding median error. We can conclude that gammatonegram segmentation is, indeed, beneficial to sound source localisation as well; our proposed method (\textit{cf.} Table~\ref{table:exp_reg}) provides $\tilde{E}_{SeC} = 7.5\degree$, in contrast to $11.02\degree$\rv{,} $16.98\degree$ \rv{and $10.1\degree$}, respectively (corresponding to $31\%$\rv{,} $56\%$\rv{, and $25\%$} improvement).

All the experiments considered so far, rely on a random split of the dataset into training and test sets. This scenario, however, is particularly challenging, and does not adhere faithfully to the reality, where the testing frames will be part of the same data stream and additional processing can be applied. In this last experiment, we build an additional test dataset, consisting of $600$ consecutive audio frames and apply median filters of different orders to the DoA estimates to remove potential outliers. Results are shown in Fig. \ref{fig:reg_med}. We observe that the system becomes extremely reliable: we provide, for instance, estimates with a median absolute error of $2.5\degree$ when employing audio frames of $2.5s$, which is an acceptable time frame in our scenario, as our system works properly also at really low SNRs, which implies that the sound source is still at a considerable distance from the microphones.

\subsection{\rrv{Comparison with Related Work and Discussion}}
\label{subsec:discussion}
\rv{In order to also have a means of comparing the performance of our system  to other related works, Table \ref{tab:comparison} shows the accuracy achieved by those works, together with some of the corresponding experimental settings, as reported by the authors in the original publications. Even though a direct comparison is not feasible, all works rely on different datasets and experimental conditions, we can observe that, operating on a more extensive dataset, in much harsher noisy scenarios, and while carrying out multi-class classification rather than detection (as almost all the other works used in the comparison do), the performance reached by our framework is comparable, when not greater, than what achieved by other methods.}
\rrv{As an additional measure of comparison, we adapt the methods proposed in \cite{Salamon:UrbanSound:ACMMM:14} to operate on our dataset and analyse their performance. Specifically, we investigate the behaviour of a system relying on Support Vector Machines (\emph{SVMs}) and one relying on Random Forests (\emph{RFs}), both acting on MFCCs-based feature vectors, adjusted to execute the three-class classification task of our framework. Figure \ref{fig:class_snr} reports the results of this comparison. We can observe that our framework  outperforms both the \emph{SVM} system and the \emph{RF} one, and that this difference increases with the amount of noise. Indeed, while the performance of our framework are only mildly affected by the lowering of the SNR, the recognition rate obtained by the SVM and RF drop dramatically at lower SNRs til reaching less than $50\%$ when $SNR=-35dB$.}

\begin{table}[]
\resizebox{\columnwidth}{!}{%
\renewcommand{\arraystretch}{1}
\begin{tabular}{c|c|c|c|c|c}
\toprule
\textbf{Approach} & \textbf{ACC S} & \textbf{ACC H} & \textbf{SNR} & \textbf{Dataset}  & \textbf{Loc.} \\ 
&{[}\%{]}\textbf{}&{[}\%{]}&\textbf{\textbf{{[}}dB{]}}&\textbf{\textbf{{[}}\# samples{]}} & \\ \midrule
SVMs \& RFs \cite{Salamon:UrbanSound:ACMMM:14}           & 75\%               & 75\%             & N/S         & $\sim$9K (1K per class)     & -                        \\ \hline
Part-based Models \cite{schroder2013automatic}         & 86\%               &  -          & [-20,20]         &  $\sim$1K    &  -   \\       
 \hline
    Predictive Coding \cite{anacur2019detecting}           & 94.7\%               & -            & -         & $\sim$100     & -                         \\ \hline
       Fourier Decomposition \cite{fatimah2020automatic}         & 98.5\%               & -           & N/S         & N/S     & -                             \\ \hline
         Spectral Analysis  \cite{ebizuka2019detecting}     & 100\%                & -           & 20         & N/S      & -                              \\ \hline
          SVMs \cite{carmel2017detection}      & 98\%                 & -           & N/S         & $\sim$2K        & -                       \\ \hline
        Gaussian Mixture Models \cite{banerjee2013participatory}        & -              & 88.3\%             & N/S         & 800     & -           \\ \hline
        Prior: KNN~\cite{marchegiani2017leveraging}         & 81\%               & 79\%            & N/S        &  $\sim$10K (3.3K per class)      &  - \\ \hline
  This work         & 98\%               & 90\%            & [-40, 10]        &  30K      &  \checkmark   
                                          \\ \bottomrule
\end{tabular}}
\caption{\rv{Table shows an indirect comparison with other works in alarming sound detection (\textit{Approach}), reporting the respective average accuracy obtained in the detection of sirens (\textit{ACC S}) and horns (\textit{ACC H}), the noise conditions (\textit{SNR}), the size of the dataset utilised in the experiments, and whether source localisation (\textit{Loc.}) was performed. `-' stands for `not applicable', while `N/S' stands for `not specified'.}}
\label{tab:comparison}
\vspace{-5.5mm}
\end{table}

\begin{figure*}%
\centering
\begin{subfigure}{.45\columnwidth}
\includegraphics[width=\columnwidth]{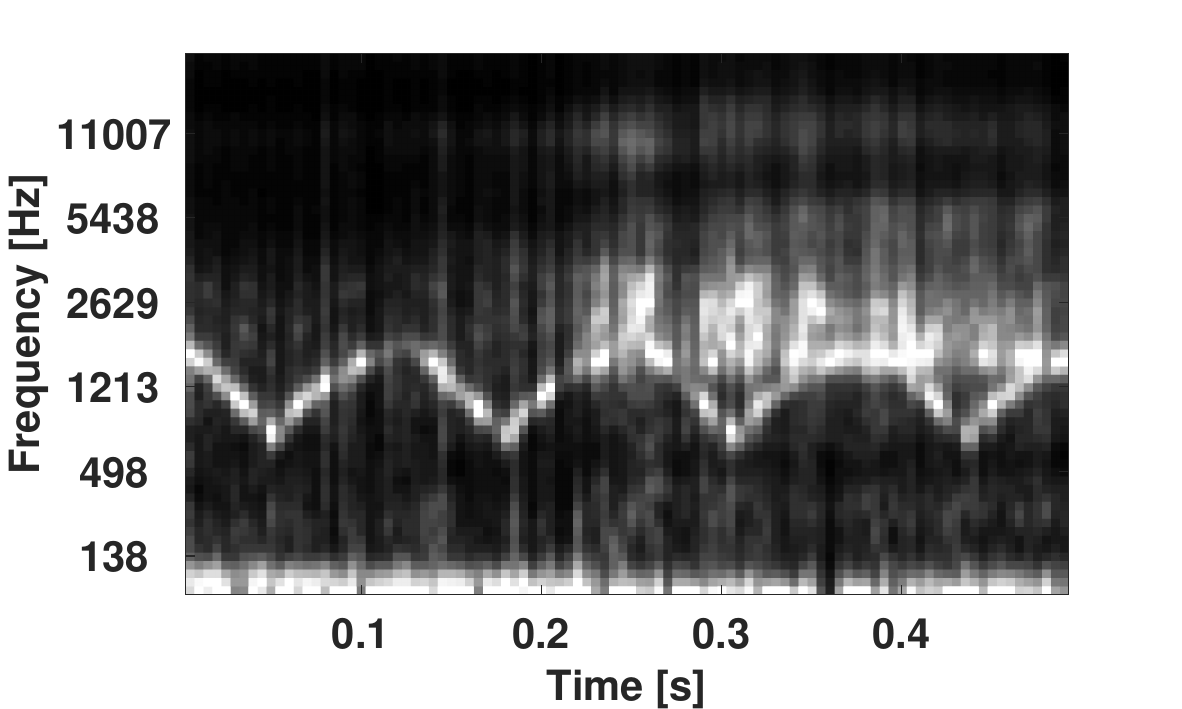}%
\caption{Noisy Signal}%
\label{subfig:noisy}%
\end{subfigure}\hfill%
\begin{subfigure}{.45\columnwidth}
\includegraphics[width=\columnwidth]{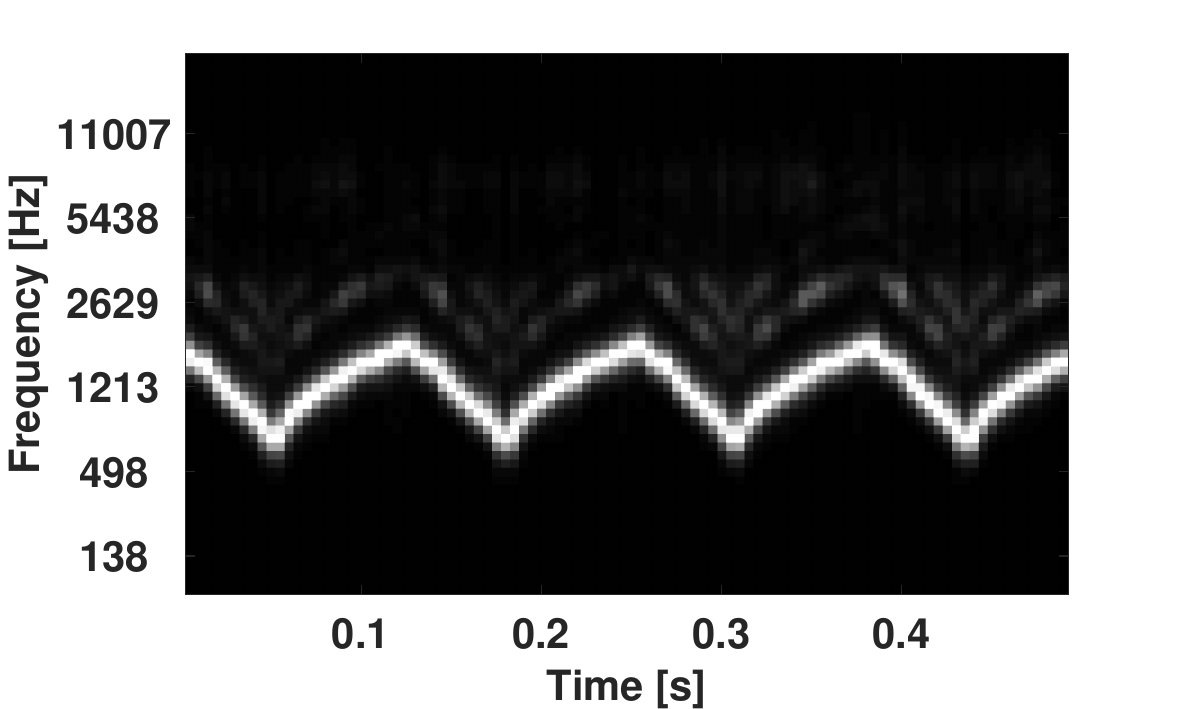}%
\caption{Ground Truth}%
\label{subfig:gt}%
\end{subfigure}\hfill%
\begin{subfigure}{.45\columnwidth}
\includegraphics[width=\columnwidth]{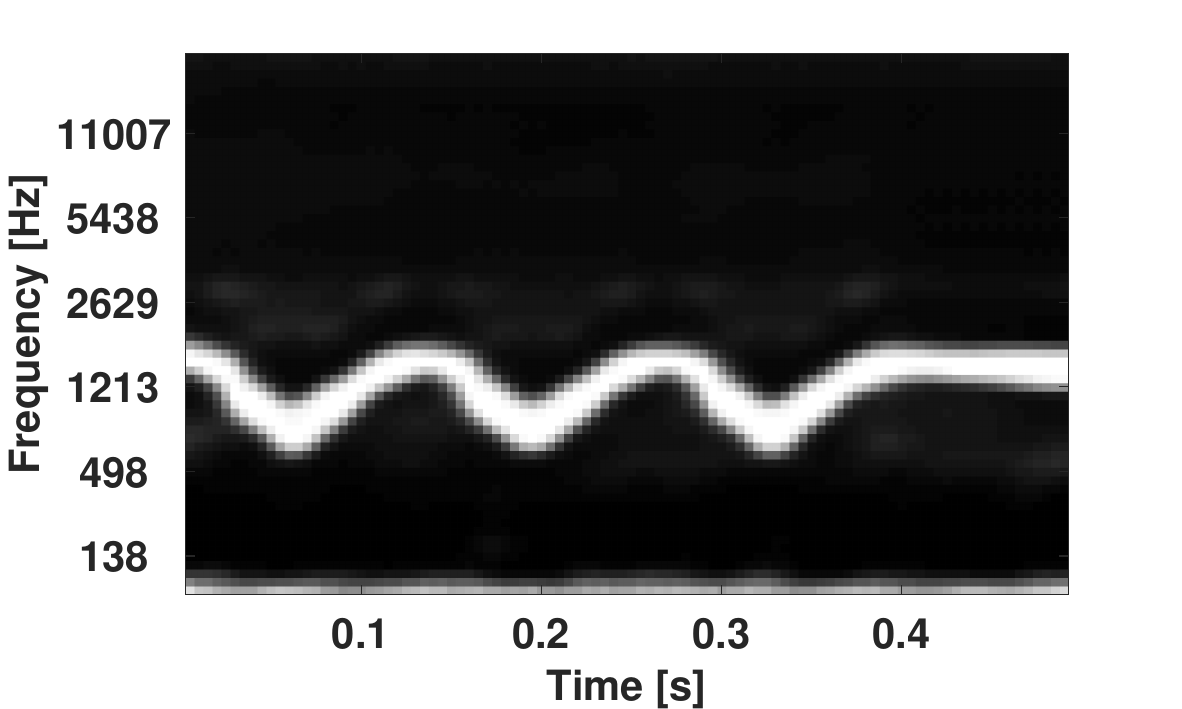}%
\caption{RELU-based \emph{SeC}}%
\label{subfig:relu}%
\end{subfigure}\hfill%
\begin{subfigure}{.45\columnwidth}
\includegraphics[width=\columnwidth]{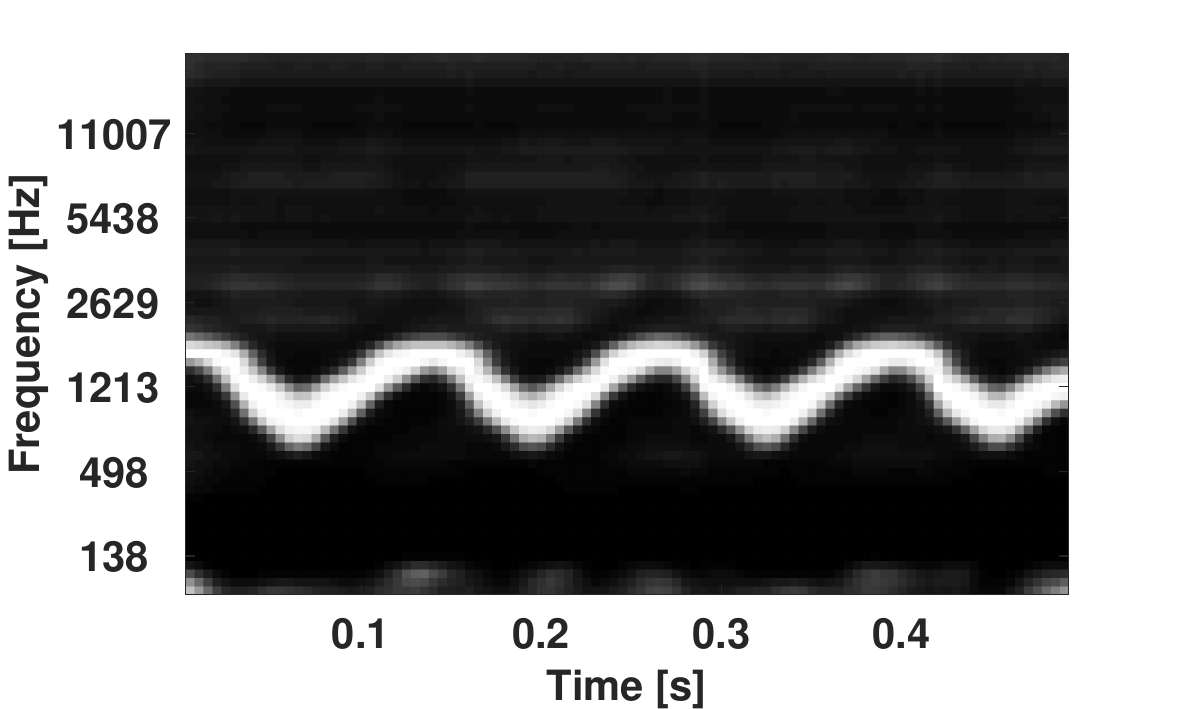}%
\caption{ELU-based \emph{SeC}}%
\label{subfig:elu}%
\end{subfigure}%
\caption{\rv{Example of segmentation mask obtained by employing the \emph{SeC} network with RELUs (c) and with ELUs (d). The figures also reports the initial noisy signal (a) and the corresponding ground truth segmentation mask (b).}}
	\label{fig:relus}
	\vspace{-2.5mm}
\end{figure*}

\rv{We would also like to draw to the attention of the reader that in all the deep learning architectures analysed in this work, all convolutions occur with Exponential Linear Unit (ELU). The reason for this choice is to be attributed to some preliminary experiments we have performed and which demonstrated that the use of ELUs in these scenarios led to higher performance with respect to other activation functions. In particular, in the case of Rectified Linear Units (ReLUs), while no significant differences in the performance were observed in the event classification task (\textit{cf.} \emph{SeC} network), the regression task (\textit{cf.} \emph{SL} network) was characterised by a median error of $10 \degree$, compared to the $\tilde{E}_{SeC} = 7.5\degree$, obtained with the use of ELUs. Further investigations seem to suggest that the reason behind this behaviour is due to a better segmentation mask generated by the \emph{SeC} network when using ELUs. One example is shown in Fig.~\ref{fig:relus} where it is possible to observe that the segmentation mask generated by employing RELUs is less resilient to noise.}

\rv{We did not directly analyse the real-time performance of the system, as we do not anticipate that this would be a challenge. Indeed, we do not foresee additional delays with respect to the offline experiments we carried out, considering that the system runs roughly at 2~Hz (frequency dictated by the length of the audio frames we employ). Indeed, while the training is time consuming, the execution time of sensing, gammatonegram creation and inference (through the deep network) is insignificant with respect to the 2~Hz (0.5s window).}

\section{CONCLUSIONS}

In this paper we proposed a framework to detect alerting sound events in  urban environments, and localise the respective sound source. As traffic scenarios are characterised by copious, unstructured and unpredictable noise, we proposed a new denoising method based on semantic segmentation of the stereo gammatonegram of the signals in a multi-task learning scheme, to simultaneously recover the original clean sound, and identify its nature. The direction of arrival of the sound is, then, obtained by training a CNN with the cross-gammatongrams of the denoised signals. Our experimental evaluation, which includes challenging scenarios characterised by extremely low SNRs (\textit{$-40 dB  \leq SNR \leq 10 dB$}), showed an average classification rate of $94 \%$, and a median absolute error of $7.5\degree$ when operating on audio frames of $0.5s$, and of $2.5\degree$ when operating with a median filter on longer frames of  $2.5s$. \rv{Further work can investigate the detection of audio events against other noise sources, and the occurrence of simultaneous acoustic alarms of a different nature.}

\bibliographystyle{IEEEtran}
\bibliography{references}

\begin{thebibliography}{10}
\providecommand{\url}[1]{#1}
\csname url@samestyle\endcsname
\providecommand{\newblock}{\relax}
\providecommand{\bibinfo}[2]{#2}
\providecommand{\BIBentrySTDinterwordspacing}{\spaceskip=0pt\relax}
\providecommand{\BIBentryALTinterwordstretchfactor}{4}
\providecommand{\BIBentryALTinterwordspacing}{\spaceskip=\fontdimen2\font plus
\BIBentryALTinterwordstretchfactor\fontdimen3\font minus
  \fontdimen4\font\relax}
\providecommand{\BIBforeignlanguage}[2]{{%
\expandafter\ifx\csname l@#1\endcsname\relax
\typeout{** WARNING: IEEEtran.bst: No hyphenation pattern has been}%
\typeout{** loaded for the language `#1'. Using the pattern for}%
\typeout{** the default language instead.}%
\else
\language=\csname l@#1\endcsname
\fi
#2}}
\providecommand{\BIBdecl}{\relax}
\BIBdecl

\bibitem{tatoglu2019self}
A.~Tatoglu and E.~King, ``Self driving car path planning modification with
  respect to rapid emergency vehicle detection,'' in \emph{INTER-NOISE and
  NOISE-CON Congress and Conference Proceedings}, vol. 259, no.~3.\hskip 1em
  plus 0.5em minus 0.4em\relax Institute of Noise Control Engineering, 2019,
  pp. 6282--6289.

\bibitem{hersh2010assistive}
M.~Hersh and M.~A. Johnson, \emph{Assistive technology for visually impaired
  and blind people}.\hskip 1em plus 0.5em minus 0.4em\relax London, U.K.:
  Springer, 2010.

\bibitem{marchegiani2017leveraging}
L.~Marchegiani and I.~Posner, ``Leveraging the urban soundscape: Auditory
  perception for smart vehicles,'' in \emph{Robotics and Automation (ICRA),
  2017 IEEE Int. Conf. on}, 2017, pp. 6547--6554.

\bibitem{long2015fully}
J.~Long, E.~Shelhamer, and T.~Darrell, ``Fully convolutional networks for
  semantic segmentation,'' in \emph{Proceedings of the IEEE conference on
  computer vision and pattern recognition}, 2015, pp. 3431--3440.

\bibitem{ronneberger2015u}
O.~Ronneberger, P.~Fischer, and T.~Brox, ``U-net: Convolutional networks for
  biomedical image segmentation,'' in \emph{Int. Conf. on Medical image
  computing and computer-assisted intervention}, 2015, pp. 234--241.

\bibitem{argentieri2015survey}
S.~Argentieri, P.~Dan{\`e}s, and P.~Sou{\`e}res, ``A survey on sound source
  localization in robotics: From binaural to array processing methods,''
  \emph{Computer Speech \& Language}, vol.~34, no.~1, pp. 87--112, 2015.

\bibitem{parry2020pressure}
J.~A. Parry, K.~V. Horoshenkov, and D.~P. Williams, ``Pressure ratio and phase
  difference in a two-microphone system under uncertain outdoor sound
  propagation conditions,'' \emph{Applied Acoustics}, vol. 170, 2020.

\bibitem{fazenda2009acoustic}
B.~Fazenda, H.~Atmoko, F.~Gu, L.~Guan, and A.~Ball, ``Acoustic based safety
  emergency vehicle detection for intelligent transport systems,'' in
  \emph{IEEE ICROS-SICE Int. Joint Conference 2009}, 2009.

\bibitem{meucci2008real}
F.~Meucci, L.~Pierucci, E.~Del~Re, L.~Lastrucci, and P.~Desii, ``A real-time
  siren detector to improve safety of guide in traffic environment,'' in
  \emph{Signal Processing Conference, IEEE 16th European}, 2008, pp. 1--5.

\bibitem{schroder2013automatic}
J.~Schr{\"o}der, S.~Goetze, V.~Grutzmacher, and J.~Anem{\"u}ller, ``Automatic
  acoustic siren detection in traffic noise by part-based models.'' in
  \emph{ICASSP}, 2013, pp. 493--497.

\bibitem{ntalampiras2009adaptive}
S.~Ntalampiras, I.~Potamitis, and N.~Fakotakis, ``An adaptive framework for
  acoustic monitoring of potential hazards,'' \emph{EURASIP Journal on Audio,
  Speech, and Music Processing}, vol. 2009, p.~13, 2009.

\bibitem{Salamon:UrbanSound:ACMMM:14}
J.~Salamon, C.~Jacoby, and J.~P. Bello, ``A dataset and taxonomy for urban
  sound research,'' in \emph{22st {ACM} Int. Conf. on Multimedia
  ({ACM-MM'14})}, Orlando, FL, USA, Nov. 2014.

\bibitem{carmel2017detection}
D.~Carmel, A.~Yeshurun, and Y.~Moshe, ``Detection of alarm sounds in noisy
  environments,'' in \emph{2017 25th European Signal Processing Conference
  (EUSIPCO)}.\hskip 1em plus 0.5em minus 0.4em\relax IEEE, 2017, pp.
  1839--1843.

\bibitem{banerjee2013participatory}
R.~Banerjee, A.~Sinha, and A.~Saha, ``Participatory sensing based traffic
  condition monitoring using horn detection,'' in \emph{Proceedings of the 28th
  annual ACM symposium on applied computing}, 2013, pp. 567--569.

\bibitem{ebizuka2019detecting}
Y.~Ebizuka, S.~Kato, and M.~Itami, ``Detecting approach of emergency vehicles
  using siren sound processing,'' in \emph{2019 IEEE Intelligent Transportation
  Systems Conference (ITSC)}, 2019, pp. 4431--4436.

\bibitem{widrow1985adaptive}
B.~Widrow and S.~D. Stearns, ``Adaptive signal processing,'' \emph{Englewood
  Cliffs, NJ, Prentice-Hall, Inc., 1985, 491 p.}, vol.~1, 1985.

\bibitem{anacur2019detecting}
C.~A. Anacur and R.~Saracoglu, ``Detecting of warning sounds in the traffic
  using linear predictive coding,'' \emph{Int. Journal of Intelligent Systems
  and Applications in Engineering}, vol.~7, no.~4, pp. 195--200, 2019.

\bibitem{fatimah2020automatic}
B.~Fatimah, A.~Preethi, V.~Hrushikesh, A.~Singh, and H.~R. Kotion, ``An
  automatic siren detection algorithm using fourier decomposition method and
  mfcc,'' in \emph{2020 11th IEEE Int. Conf. on Computing, Communication and
  Networking Technologies (ICCCNT)}, 2020, pp. 1--6.

\bibitem{badrinarayanan2015segnet}
V.~Badrinarayanan, A.~Kendall, and R.~Cipolla, ``Segnet: A deep convolutional
  encoder-decoder architecture for image segmentation,'' \emph{arXiv preprint
  arXiv:1511.00561}, 2015.

\bibitem{papandreou2015weakly}
G.~Papandreou, L.-C. Chen, K.~Murphy, and A.~L. Yuille, ``Weakly-and
  semi-supervised learning of a dcnn for semantic image segmentation,''
  \emph{arXiv preprint arXiv:1502.02734}, 2015.

\bibitem{deshpande2001classification}
H.~Deshpande, R.~Singh, and U.~Nam, ``Classification of music signals in the
  visual domain,'' in \emph{Proceedings of the COST-G6 Conference on Digital
  Audio Effects}, 2001, pp. 1--4.

\bibitem{dutta2007text}
T.~Dutta, ``Text dependent speaker identification based on spectrograms,''
  \emph{Proceedings of Image and vision computing}, pp. 238--243, 2007.

\bibitem{towhid2017spectrogram}
M.~S. Towhid and M.~M. Rahman, ``Spectrogram segmentation for bird species
  classification based on temporal continuity,'' in \emph{20th IEEE Int. Conf.
  of Computer and Information Technology (ICCIT)}, 2017, pp. 1--4.

\bibitem{fu1981survey}
K.-S. Fu and J.~Mui, ``A survey on image segmentation,'' \emph{Pattern
  recognition}, vol.~13, no.~1, pp. 3--16, 1981.

\bibitem{mulimani2019segmentation}
M.~Mulimani and S.~G. Koolagudi, ``Segmentation and characterization of
  acoustic event spectrograms using singular value decomposition,''
  \emph{Expert Systems with Applications}, vol. 120, pp. 413--425, 2019.

\bibitem{rudzyn2007real}
B.~Rudzyn, W.~Kadous, and C.~Sammut, ``Real time robot audition system
  incorporating both 3d sound source localisation and voice characterisation,''
  in \emph{Robotics and Automation, 2007 IEEE Int. Conf. on}, 2007, pp.
  4733--4738.

\bibitem{marchegiani2009multimodal}
L.~Marchegiani, F.~Pirri, and M.~Pizzoli, ``Multimodal speaker recognition in a
  conversation scenario,'' in \emph{Int. Conf. on Computer Vision Systems},
  2009, pp. 11--20.

\bibitem{he2017deep}
W.~He, P.~Motlicek, and J.-M. Odobez, ``Deep neural networks for multiple
  speaker detection and localization,'' \emph{arXiv preprint arXiv:1711.11565},
  2017.

\bibitem{ma2017exploiting}
N.~Ma, T.~May, and G.~J. Brown, ``Exploiting deep neural networks and head
  movements for robust binaural localization of multiple sources in reverberant
  environments,'' \emph{IEEE/ACM Trans. Audio, Speech, Language Process},
  vol.~25, no.~12, pp. 2444--2453, 2017.

\bibitem{tse2019no}
T.~H.~E. Tse, D.~De~Martini, and L.~Marchegiani, ``No need to scream: Robust
  sound-based speaker localisation in challenging scenarios,'' in \emph{Int.
  Conf. on Social Robotics}, 2019, pp. 176--185.

\bibitem{vera2018towards}
J.~M. Vera-Diaz, D.~Pizarro, and J.~Macias-Guarasa, ``Towards end-to-end
  acoustic localization using deep learning: From audio signals to source
  position coordinates,'' \emph{Sensors}, vol.~18, no.~10, p. 3418, 2018.

\bibitem{he2019adaptation}
W.~He, P.~Motlicek, and J.-M. Odobez, ``Adaptation of multiple sound source
  localization neural networks with weak supervision and domain-adversarial
  training,'' in \emph{ICASSP IEEE Int. Conf. on Acoustics, Speech and Signal
  Processing (ICASSP)}, 2019, pp. 770--774.

\bibitem{zhang2018new}
X.~Zhang, H.~Sun, S.~Wang, and J.~Xu, ``A new regional localization method for
  indoor sound source based on convolutional neural networks,'' \emph{IEEE
  Access}, vol.~6, pp. 72\,073--72\,082, 2018.

\bibitem{zhuang2008feature}
X.~Zhuang, X.~Zhou, T.~S. Huang, and M.~Hasegawa-Johnson, ``Feature analysis
  and selection for acoustic event detection,'' in \emph{2008 IEEE Int. Conf.
  on Acoustics, Speech and Signal Processing}, 2008, pp. 17--20.

\bibitem{huang2016autonomous}
S.-W. Huang, N.~Taniguchi, A.-T. Hsiao, C.-F. Huang, E.~Chen, C.-L. Ting, and
  J.~Guo, ``Autonomous underwater vehicle localization using ocean tomography
  sensor nodes,'' in \emph{OCEANS 2016 MTS/IEEE Monterey}, 2016, pp. 1--5.

\bibitem{rypkema2018closed}
N.~R. Rypkema, E.~M. Fischel, and H.~Schmidt, ``Closed-loop single-beacon
  passive acoustic navigation for low-cost autonomous underwater vehicles,'' in
  \emph{2018 IEEE/RSJ Int. Conf. on Intelligent Robots and Systems (IROS)},
  2018, pp. 641--648.

\bibitem{mennitt2010multiple}
D.~Mennitt and M.~Johnson, ``Multiple-array passive acoustic source
  localization in urban environments,'' \emph{The Journal of the Acoustical
  Society of America}, vol. 127, no.~5, pp. 2932--2942, 2010.

\bibitem{faraji2019sound}
M.~M. Faraji, S.~B. Shouraki, E.~Iranmehr, and B.~Linares-Barranco, ``Sound
  source localization in wide-range outdoor environment using distributed
  sensor network,'' \emph{IEEE Sensors Journal}, vol.~20, no.~4, pp.
  2234--2246, 2019.

\bibitem{ding2004adaptive}
H.~Ding, J.~Lu, X.~Qiu, and B.~Xu, ``An adaptive speech enhancement method for
  siren noise cancellation,'' \emph{Applied Acoustics}, vol.~65, no.~4, pp.
  385--399, 2004.

\bibitem{wagner2000guide}
R.~P. Wagner, ``Guide to test methods, performance requirements, and
  installation practices for electronic sirens used on law enforcement
  vehicles,'' National Institute of Justice (NIJ), U.S. Department of Justice.
  Washington, D.C., NIJ Guide 500–00, Tech. Rep., 2000.

\bibitem{chakrabarty2016abnormal}
D.~Chakrabarty and M.~Elhilali, ``Abnormal sound event detection using temporal
  trajectories mixtures,'' in \emph{IEEE Int. Conf. on Acoustics, Speech and
  Signal Processing (ICASSP)}, 2016, pp. 216--220.

\bibitem{lyon2010history}
R.~F. Lyon, A.~G. Katsiamis, and E.~M. Drakakis, ``History and future of
  auditory filter models,'' in \emph{Proceedings of 2010 IEEE Int. Symposium on
  Circuits and Systems}, 2010, pp. 3809--3812.

\bibitem{holdsworth1988implementing}
J.~Holdsworth, I.~Nimmo-Smith, R.~Patterson, and P.~Rice, ``Implementing a
  gammatone filter bank,'' \emph{Annex C of the SVOS Final Report: Part A: The
  Auditory Filterbank}, vol.~1, pp. 1--5, 1988.

\bibitem{toshio1995optimal}
I.~Toshio, ``An optimal auditory filter,'' in \emph{Applications of Signal
  Processing to Audio and Acoustics, IEEE ASSP Workshop}, 1995, pp. 198--201.

\bibitem{glasberg1990derivation}
B.~R. Glasberg and B.~C. Moore, ``Derivation of auditory filter shapes from
  notched-noise data,'' \emph{Hearing Res.}, vol.~47, no.~1, pp. 103--138,
  1990.

\bibitem{collobert2008unified}
R.~Collobert and J.~Weston, ``A unified architecture for natural language
  processing: Deep neural networks with multitask learning,'' in \emph{25th
  Int. Conf. on Machine learning}, 2008, pp. 160--167.

\bibitem{jin2008neural}
F.~Jin and S.~Sun, ``Neural network multitask learning for traffic flow
  forecasting,'' in \emph{Neural Networks, 2008. IJCNN 2008.(IEEE World
  Congress on Computational Intelligence). IEEE Int. Joint Conference on},
  2008, pp. 1897--1901.

\bibitem{caruana1998multitask}
R.~Caruana, ``Multitask learning: A knowledge-based source of inductive bias.''
  in \emph{Proc. Int. Conf. Machine Learning}, 1993, pp. 41--48.

\bibitem{ClevertUH15}
D.~Clevert, T.~Unterthiner, and S.~Hochreiter, ``Fast and accurate deep network
  learning by exponential linear units (elus),'' \emph{CoRR}, vol.
  abs/1511.07289, 2015.

\bibitem{chan2016listen}
W.~Chan, N.~Jaitly, Q.~Le, and O.~Vinyals, ``Listen, attend and spell: A neural
  network for large vocabulary conversational speech recognition,'' in
  \emph{Acoustics, Speech and Signal Processing (ICASSP), 2016 IEEE Int. Conf.
  on}, 2016, pp. 4960--4964.

\bibitem{marchegiani2015cross}
L.~Marchegiani and X.~Fafoutis, ``On cross-language consonant identification in
  second language noise,'' \emph{The Journal of the Acoustical Society of
  America}, vol. 138, no.~4, pp. 2206--2209, 2015.

\bibitem{noda2015sound}
K.~Noda, N.~Hashimoto, K.~Nakadai, and T.~Ogata, ``Sound source separation for
  robot audition using deep learning,'' in \emph{Humanoid Robots (Humanoids),
  2015 IEEE-RAS 15th Int. Conf. on}, 2015, pp. 389--394.

\bibitem{kingma2014adam}
D.~Kingma and J.~Ba, ``Adam: A method for stochastic optimization,'' \emph{3rd
  Int. Conf. for Learning Representations (ICLR)}, pp. 1--15, 2015.

\bibitem{hinton2012improving}
G.~E. Hinton, N.~Srivastava, A.~Krizhevsky, I.~Sutskever, and R.~R.
  Salakhutdinov, ``Improving neural networks by preventing co-adaptation of
  feature detectors,'' \emph{arXiv preprint arXiv:1207.0580}, 2012.

\bibitem{tensorflow2015-whitepaper}
\BIBentryALTinterwordspacing
M.~A. et~al., ``{TensorFlow}: Large-scale machine learning on heterogeneous
  systems,'' 2015, software available from tensorflow.org. [Online]. Available:
  \url{http://tensorflow.org/}
\BIBentrySTDinterwordspacing

\bibitem{deng2014robust}
S.~Deng, J.~Han, C.~Zhang, T.~Zheng, and G.~Zheng, ``Robust minimum statistics
  project coefficients feature for acoustic environment recognition,'' in
  \emph{Acoustics, Speech and Signal Processing (ICASSP), 2014 IEEE Int. Conf.
  on}, 2014, pp. 8232--8236.

\bibitem{takahashi2017aenet}
N.~Takahashi, M.~Gygli, and L.~Van~Gool, ``Aenet: Learning deep audio features
  for video analysis,'' \emph{arXiv preprint arXiv:1701.00599}, 2017.

\end{thebibliography}

\vspace*{-2\baselineskip}
\begin{IEEEbiography}[{\includegraphics[width=1in,height=1.25in,clip,keepaspectratio]{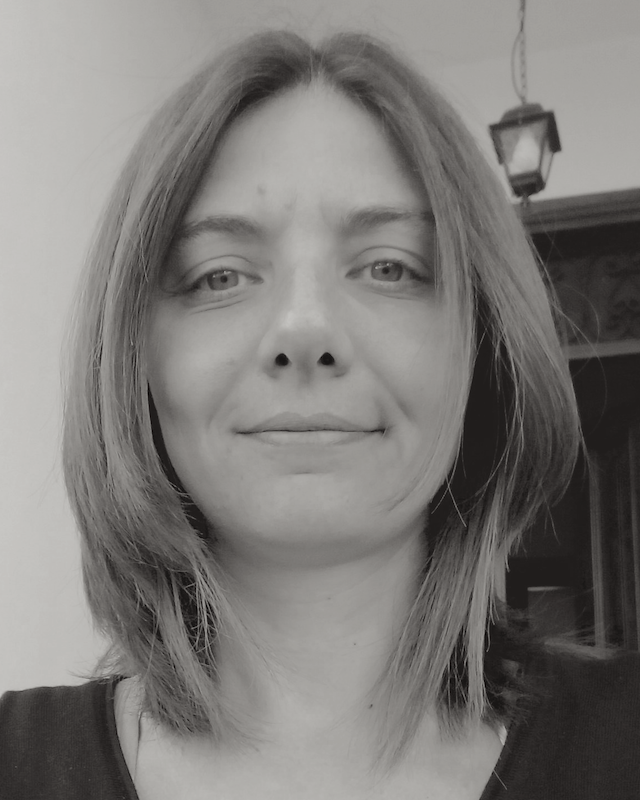}}]{Letizia Marchegiani}
Letizia Marchegiani (Memeber, IEEE) received the B.Sc., M.Sc., and Ph.D. degrees in computer engineering from the Sapienza University of Rome,  Italy, in 2005, 2008, and 2012, respectively. From
2014 to 2018, she was a researcher at the
University of Oxford (UK), where she was a member
of the Oxford Robotics Institute (ORI). 
Since 2019 she is an Assistant Professor in Robotics with the Department of Electronic Systems of the Aalborg University (Denmark). Her research
interests primarily lie in the areas of signal processing, machine learning, and their application to robotics, autonomous systems, cognitive modelling, intelligent transportation, and digital healthcare.
\end{IEEEbiography}
\vspace*{-3\baselineskip}
\begin{IEEEbiography}[{\includegraphics[width=1in,height=1.25in,clip,keepaspectratio]{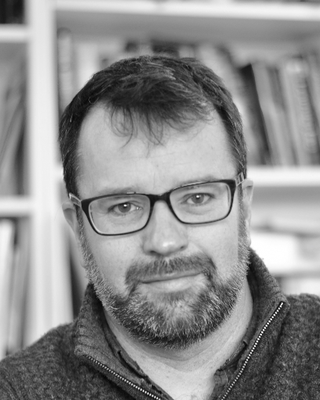}}]{Paul Newman}
Paul Newman (Fellow, IEEE) received the M.Eng. degree in engineering science from University of Oxford, Oxford, U.K., in 1995 and the Ph.D. degree in autonomous navigation from the Australian Center for Field Robotics, University of Sydney, Sydney, Australia. He is currently a BP Professor of information engineering with the Department of Engineering Science, University of Oxford, and and Fellow of Keble College. He is also Director of the Oxford Robotics Institute, and Founder of of Oxbotica Lt. 
He was elected Fellow of the Royal Academy of Engineering and Fellow of the IEEE in 2014 both with citations for outstanding contributions to robot navigation. 
He serves on the UK Department for Transport's Scientific Advisory Council, and was an architect of the UK’s Robotics and Autonomous Systems Strategy. 
His research interests include autonomous navigation, especially over large spatial and temporal scale.
\end{IEEEbiography}





\end{document}